\documentclass[sigconf]{acmart}
\AtBeginDocument{%
  \providecommand\BibTeX{{%
    \normalfont B\kern-0.5em{\scshape i\kern-0.25em b}\kern-0.8em\TeX}}}

\acmISBN{978-1-4503-XXXX-X/18/06}

\usepackage{float}
\usepackage{newfloat}
\usepackage{multirow}
\usepackage{makecell}
\usepackage{xcolor}         
\usepackage{colortbl}
\usepackage{dsfont}
\usepackage{amsmath}
\usepackage{enumitem}

\usepackage[normalem]{ulem}
\useunder{\uline}{\ul}{}

\definecolor{myyellow}{rgb}{1,1, 0.6}
\definecolor{myorange}{rgb}{1, 0.8, 0.6}
\definecolor{myred}{rgb}{1, 0.6, 0.6}
\definecolor{second}{HTML}{FFDAB9}
\definecolor{best}{HTML}{FFC1C1}
\usepackage{multirow}
\usepackage{colortbl}
\usepackage[normalem]{ulem}
\usepackage{bbding}
\useunder{\uline}{\ul}{}
\usepackage{stackrel}
\usepackage{enumitem}
\usepackage{algpseudocode}
\usepackage[ruled,lined]{algorithm2e}
\usepackage[normalem]{ulem}
\usepackage{tcolorbox}
\usepackage{balance}
\usepackage{hhline} 

\DeclareMathOperator*{\argminA}{arg\,min} 

\author{Zheqi Lv}
\affiliation{%
  \institution{Zhejiang University}
  \city{Hangzhou}
  \country{China}}
\email{zheqilv@zju.edu.cn}

\author{Tianyu Zhan}
\affiliation{%
  \institution{Zhejiang University}
  \city{Hangzhou}
  \country{China}}
\email{yuzt@zju.edu.cn}

\author{Wenjie Wang}
\affiliation{%
  \institution{National University of Singapore}
  \city{Singapore}
  \country{Singapore}}
\authornote{Corresponding authors.}
\email{wenjiewang96@gmail.com}

\author{Xinyu Lin}
\affiliation{%
  \institution{National University of Singapore}
  \city{Singapore}
  \country{Singapore}}
\email{xylin1028@gmail.com}

\author{Shengyu Zhang}
\affiliation{%
  \institution{Zhejiang University}
  \city{Hangzhou}
  \country{China}}
\authornotemark[1]
\email{sy_zhang@zju.edu.cn}

\author{Wenqiao Zhang}
\affiliation{%
  \institution{Zhejiang University}
  \city{Hangzhou}
  \country{China}}
\email{wenqiaozhang@zju.edu.cn}

\author{Jiwei Li}
\affiliation{%
  \institution{Zhejiang University}
  \city{Hangzhou}
  \country{China}}
\email{jiwei_li@zju.edu.cn}

\author{Kun Kuang}
\affiliation{%
  \institution{Zhejiang University}
  \city{Hangzhou}
  \country{China}}
\authornotemark[1]
\email{kunkuang@zju.edu.cn}

\author{Fei Wu}
\affiliation{%
  \institution{Zhejiang University}
  \city{Hangzhou}
  \country{China}}
\email{wufei@zju.edu.cn}

\copyrightyear{2025}
\acmYear{2025}
\setcopyright{acmlicensed}\acmConference[KDD '25]{Proceedings of the 31st ACM SIGKDD Conference on Knowledge Discovery and Data Mining V.1}{August 3--7, 2025}{Toronto, ON, Canada}
\acmBooktitle{Proceedings of the 31st ACM SIGKDD Conference on Knowledge Discovery and Data Mining V.1 (KDD '25), August 3--7, 2025, Toronto, ON, Canada}
\acmDOI{10.1145/3690624.3709335}
\acmISBN{979-8-4007-1245-6/25/08}

\begin{document}

\title{Collaboration of Large Language Models and Small Recommendation Models for Device-Cloud Recommendation}

\renewcommand{\shortauthors}{Zheqi Lv et al.}



\begin{abstract}
\label{sec:abstract}

Large Language Models (LLMs) for Recommendation (LLM4Rec) is a promising research direction that has demonstrated exceptional performance in this field. However, its inability to capture real-time user preferences greatly limits the practical application of LLM4Rec because (i) LLMs are costly to train and infer frequently, and (ii) LLMs struggle to access real-time data (its large number of parameters poses an obstacle to deployment on devices). Fortunately, small recommendation models (SRMs) can effectively supplement these shortcomings of LLM4Rec diagrams by consuming minimal resources for frequent training and inference, and by conveniently accessing real-time data on devices.

\begin{sloppypar}
In light of this, we designed the \emph{Device-Cloud \textbf{L}LM-\textbf{S}RM \textbf{C}ollaborative \textbf{Rec}ommendation Framework} (LSC4Rec) under a device-cloud collaboration setting. LSC4Rec aims to integrate the advantages of both LLMs and SRMs, as well as the benefits of cloud and edge computing, achieving a complementary synergy. We enhance the practicability of LSC4Rec by designing three strategies: collaborative training, collaborative inference, and intelligent request. 
During training, LLM generates candidate lists to enhance the ranking ability of SRM in collaborative scenarios and enables SRM to update adaptively to capture real-time user interests. 
During inference, LLM and SRM are deployed on the cloud and on the device, respectively. LLM generates candidate lists and initial ranking results based on user behavior, and SRM get reranking results based on the candidate list, with final results integrating both LLM's and SRM's scores. The device determines whether a new candidate list is needed by comparing the consistency of the LLM's and SRM's sorted lists.
Our comprehensive and extensive experimental analysis validates the effectiveness of each strategy in LSC4Rec.
\end{sloppypar}
\end{abstract}

\maketitle

\section{Introduction}
\label{sec:introduction}
Traditional sequential recommendation models, such as DIN~\cite{ref:din}, GRU4Rec~\cite{ref:gru4rec}, SASRec~\cite{ref:sasrec}, and so on, have achieved great success in both academia and industry. Here, we refer to these models that do not use large language models (LLMs) as their backbone and do not integrate user behavior into textual prompts as input for recommendation as small recommendation models (SRMs). Currently, LLMs have been explored for recommendation scenarios (LLM4Rec)~\cite{ref:llm_survey,ref:llm4rec_P5,ref:llm_rec1}, demonstrating stronger multi-task generalization capabilities and superior single-task performance, which position LLM4Rec as a highly promising research direction.

\begin{figure*}[t]
    \centering
    \includegraphics[width=0.86\textwidth]{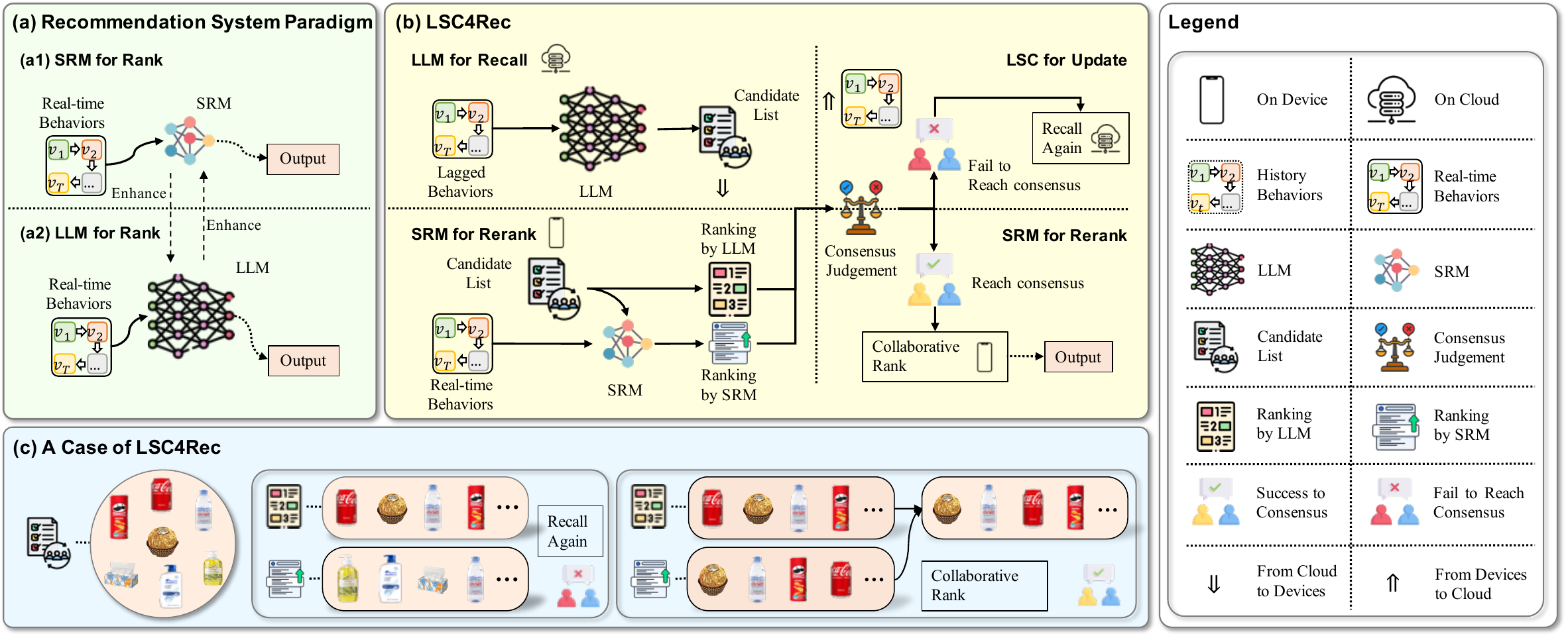}
    \vspace{-0.35cm}
    \caption{(a) respectively describes the conventional SRM4Rec and LLM4Rec diagrams. (b) is an overview of the proposed LSC4Rec, a device-cloud collaboration framework, which includes a collaborative training strategy, a collaborative inference strategy, and a collaborative update strategy. (c) is a qualitative case of our LSC4Rec framework, showing effective collaborative inference and collaborative-decision request.}
    \Description{(a) respectively describes the conventional SRM4Rec and LLM4Rec diagrams. (b) is an overview of the proposed LSC4Rec, a device-cloud collaboration framework, which includes a collaborative training strategy, a collaborative inference strategy, and a collaborative update strategy. (c) is a qualitative case of our LSC4Rec framework, showing effective collaborative inference and collaborative-decision request.}
    \label{fig:introduction}
    \vspace{-0.45cm}
\end{figure*}
However, LLMs face challenges in capturing real-time user preferences for two reasons: 
(i) \textbf{Expensive to train and inference frequently.} LLMs are built with an immense number of parameters~\cite{ref:llama,ref:llm4rec_Tallrec,ref:llm4rec_P5}, posing difficulties for the efficiency of LLM's training and inference. Moreover, the rapid increase of user behavior data exacerbates the situation, making frequent training of LLMs impractical. 
(ii) \textbf{Difficult to access real-time user behavior data.} In practical applications, user real-time behaviors are generated on devices while trainable LLMs are typically deployed on the cloud due to computational resource considerations~\cite{ref:device_cloud,ref:device_cloud_adarequest,ref:device_cloud_dccl,lv2024intelligent,lv2023duet}. Transmitting tremendous user data to the cloud in real time could cause non-trivial communication costs and latency. 

As a counterpart, SRMs could potentially supplement the shortcomings faced by LLM4Rec. (i) \textbf{SRM can be trained and inferred frequently at a significantly lower cost.} The smaller number of parameters and simpler model structure of SRM make its training and inference less resource-intensive, allowing for frequent training and inference for SRM. 
(ii) \textbf{SRM can easily access all real-time data.} The minimal use of storage and computational resources in SRM enables its deployment on devices. Therefore, accessing and handling real-time data could be much more efficient for SRM.

In this work, we propose complementing \textbf{L}LMs with \textbf{S}RMs for efficient and effective device-cloud \textbf{C}ollaborative \textbf{Rec}ommendation, namely LSC4Rec. 
The basic intuition of LSC4Rec lies in the \textit{dual optimization} of on-cloud LLMs for item candidate generation and on-device SRMs for item reranking. 
In a recommendation session, on-cloud LLMs leverage the strong user preference capturing ability to generate an item candidate list, on-device SRMs are adapted to encode the user's real-time behaviors and promptly score the items in the candidate list for real-time reranking. 
In such a device-cloud framework, LSC4Rec can reduce the training and inference frequency of on-cloud LLMs by using on-device SRMs to capture users' real-time preferences. Meanwhile, LSC4Rec can diminish the communication costs between the cloud and devices. 

However, it is non-trivial to instantiate LSC4Rec due to the following issues:
\begin{itemize}[itemsep=1pt,topsep=2pt,leftmargin=20pt]
    \item 
    How can the SRM contribute to the LSC4Rec framework? 
    Simply training the SRM and LLM separately and then deploying them in the LSC4Rec framework is empirically found ineffective, where SRM struggles to effectively rerank the item candidates from LLMs. Due to the prior knowledge of LLMs, their generated item candidates have a different distribution of negative samples from that of randomly sampled negative ones in SRM. Therefore, SRM cannot accurately distinguish the positive and negative sampled generated by the LLM during inference (Refer to empirical evidence in Table~\ref{tab:ablation_training}).  
    
    \item
    How to fully utilize the capabilities of the LLM during the inference stage? 
    In the inference stage, after the LLM generates a candidate list, only using the SRM to score items in the candidate list based on real-time data is equivalent to discarding part of the LLM's powerful generalization ability. This results in limited effectiveness of the LLM during inference, making it difficult to provide continuous benefits to the SRM's ranking.
   
    \item How to determine the shift of user preferences and update user behaviors from devices to the cloud? 
    In recommendation scenarios, user preferences are shifting over time. Therefore, it is essential to identify the user preference shifts and promptly update user behaviors from the device to the cloud, thereby obtaining a new candidate list from the LLM that better matches the user's latest preferences. 
\end{itemize}

To address the aforementioned challenges of LSC4Rec, as shown in Figure~\ref{fig:introduction}, we designed collaborative training, collaborative inference, and collaborative-decision request. 
Specifically,
\begin{itemize}[itemsep=1pt,topsep=2pt,leftmargin=20pt]
    \item Collaborative Training. Both LLM and SRM are first pre-trained based on historical data separately. Thereafter, we segment and combine items in the candidate list generated by LLM and feed this enhanced data to SRM to improve its ability to rank the candidate list retrieved by LLM. 
    At the same time, we allow SRM to undergo adaptive training to adapt to the data distribution on the device, further enhancing its ability to extract real-time user interests more effectively. 
    
    \item Collaborative Inference. LLM infers a list of candidate items and their ranking based on user behavior, which may come from a pool of millions or even billions of items. Then, SRM extracts user interests from real-time behavior to rerank the candidate items and obtain a ranking, and the reranking results of SRM are integrated with the initial ranking results of LLM through our designed score integration strategy to produce collaborative inference results. The score integration strategy includes normalizing the scores of LLM and SRM to the same range, score filtering, and weight allocation.
    
    \item Collaborative-Decision Request. It determines the necessity of requesting a new candidate list from LLM by comparing the similarity between these two sets of interests. Before outputting the results, LSC4Rec needs to compare the inconsistency of the initial ranking results based on LLM with the reranking results based on SRM. If there is low inconsistency (i.e., similar ranking results), the collaborative inference result is output. Otherwise, real-time behavior is transmitted to the LLM in the cloud to generate a new candidate list and ranking results based on short-term interests. 
\end{itemize}

In summary, our contributions include four aspects:
\begin{itemize}[itemsep=1pt,topsep=2pt,leftmargin=20pt]
\item We concentrate on a promising but under-explored research direction, specifically the challenges faced by LLMs in capturing user preferences in real-time during practical applications. 

\item We propose a novel collaborative recommendation framework called LSC4Rec. LSC4Rec effectively combines the strengths of LLMs and SRMs in a device-cloud system, paving the way for a promising direction in future research.

\item We design collaborative training, collaborative inference, and collaborative-decision request to address the three challenges of LSC4Rec and make it more practical.

\item We have conducted comprehensive and extensive experiments on various LLMs and SRMs across multiple datasets. Results show the effectiveness of LSC4Rec.
\end{itemize}
\section{Related Work}
\label{sec:related_work}
\subsection{LLM for Recommendation}
LLMs are increasingly prominent in natural language processing~\cite{lightman2023let,zhu2024model,wang2024causal,li_llm4code,zhu2024efficient,wu2024nextgpt,li2024tokenpacker,zhang2024transfr,liu2024enhancing,liu2024educating,sun2024parrot}, driving research in their application for recommendation systems~\cite{ref:instructrec,rec:recranker,ref:llm_rec_4,ref:llm4rec_P5,ref:llm4rec_Prompt_Distillation,lin2024bridging,lin2024efficient,zhang2024llasa,dai2023uncovering}. The advent of generative models like GPT transformed LLM-based recommendation into generative processes, treating recommendations as natural language tasks~\cite{ref:survey_llm4rec}. Early methods relied on prompting~\cite{ref:llm_chatrec,ref:llm_rec_2} or in-context learning~\cite{ref:llm_rec_14} but often fell short of task-specific models. Recent advancements involve fine-tuning LLMs for better alignment with recommendation tasks. P5~\cite{ref:llm4rec_P5} introduced a unified framework for fine-tuning FLAN-T5~\cite{ref:unified_trans} across various recommendation tasks. InstructRec~\cite{ref:llm4rec_understand_user_query} and TALLRec~\cite{ref:llm4rec_Tallrec} further adapted FLAN-T5 and LLaMA models, respectively, for recommendation tasks using instruction tuning. GenRec's~\cite{ref:genrec} approach involves direct instruction tuning on the LLaMA model for the generative recommendation, showcasing the evolving strategies in integrating LLMs with recommendation systems. P5~\cite{ref:llm4rec_P5} and POD~\cite{ref:llm4rec_Prompt_Distillation} are two large recommendation models based on the T5, aiming to unify the multiple recommendation tasks. They transform the recommendation task into textual format and use the text as input to the LLM to obtain recommendation results.
LLMs possess stronger generalization capabilities compared to SRMs but face challenges in terms of high costs and deployment, making it difficult to integrate real-time user preferences. The aforementioned studies have not addressed this issue. 

\subsection{Sequential Recommendation}
Sequential recommendation algorithms model users' behavior sequences and predict their next actions. They have been widely applied in scenarios such as e-commerce, short video recommendations, education, and healthcare. Classic recommendation models such as Caser~\cite{ref:caser}, GRU4Rec~\cite{ref:gru4rec}, DIN~\cite{ref:din}, and SASRec~\cite{ref:sasrec} are still commonly used today. In recent years, many studies have focused on different aspects of sequential recommendation, such as personalization \cite{lv2023duet,lv2024semantic,lv2024intelligent,sun2022response}, multimodality \cite{zhang2021mining,zhang2022latent,zhang2023mining,ji2023online}, privacy protection~\cite{liao2023ppgencdr}, user feedback~\cite{zhang2024saqrec,gao2023cirs}, cross-domain~\cite{zheng2022ddghm,liu2024learning}, cold start~\cite{liu2023joint,chen2021improving}, long-tail~\cite{zhao2023popularity}, session-based \cite{su2023enhancing}, debias~\cite{chen2022intent}, structure learning~\cite{fu2023end}, disentanglement~\cite{chen2021deep}, LLM-based~\cite{zhang2023collm,bao2023tallrec}, generative-based~\cite{zhao2024denoising,lin2024efficient}, fairness~\cite{xu2024fairrec}, etc.
Although Sequential Recommendation Models (SRMs) require fewer hardware resources and are easier to deploy on devices, they have limitations in feature extraction and generalization capabilities. These studies, however, have not addressed the deployment of SRMs on devices or the associated limitations. 

\subsection{Device-Cloud Recommendation}
In more realistic recommendation scenarios, the collaboration of device intelligence and cloud intelligence is often required~\cite{ref:device_cloud,ref:device_cloud_dccl,lv2023duet,lv2024intelligent,ref:diet,ref:device_cloud_adarequest,ref:device_cloud_rec,li2022corec,long2024diffusion}. For instance, DCCL~\cite{ref:device_cloud_dccl} and MetaController~\cite{ref:device_cloud} made early attempts at device-cloud collaborative recommendation models. DUET~\cite{lv2023duet} and IntellectReq~\cite{lv2024intelligent} explored ways to help device models overcome the generalization limitations imposed by their model size. Their specific approach involves having the device model handle personalization, while the cloud model generates the necessary parameters for the device model based on data from the device. However, when LLMs are used as cloud models, they introduce greater cloud computing loads and pose challenges in directly collaborating with SRMs. These studies do not address device-cloud collaborative recommendation strategies when LLMs serve as the cloud model.
\section{Methodology}
\label{sec:method}
\begin{figure*}
    \centering
    \includegraphics[width=0.88\textwidth]{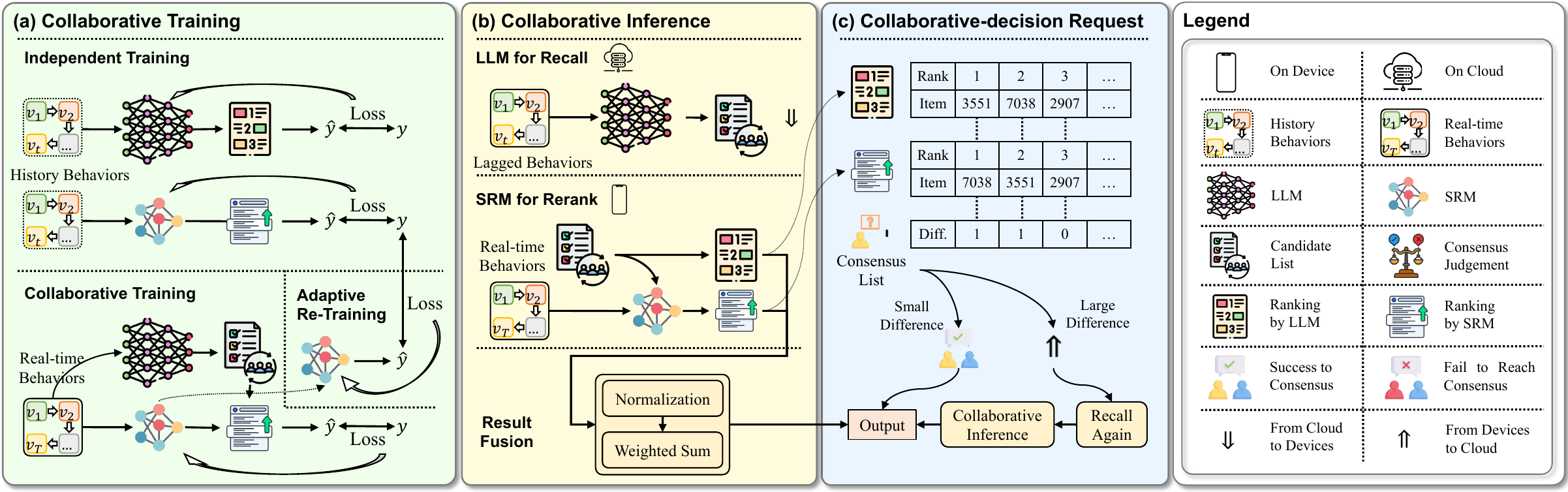}
    \vspace{-0.35cm}
    \caption{
    Overview of the LSC4Rec. (a) describes collaborative training, including independent training, cooperative training, and adaptive re-training. (b) describes collaborative inference, including cooperative inference and result fusion. (c) describes collaborative-decision request. 
    }
    \Description{
    Overview of the LSC4Rec. (a) describes collaborative training, including independent training, cooperative training, and adaptive re-training. (b) describes collaborative inference, including cooperative inference and result fusion. (c) describes collaborative-decision request. 
    }
    \label{fig:method}
    \vspace{-0.45cm}
\end{figure*}
\subsection{Problem Formulation and Notation}
\begin{sloppypar}
\subsubsection{Data.} We use $\mathcal{X}=\{u, v, s\}$ to represent a piece of data, and $\mathcal{Y}=\{y\}$ to represent the corresponding label. Specifically, $u, v, s$ represent user ID, item ID and user's click sequence respectively. 
Historical data collected for a period of time from devices is represented by $\mathcal{D}_H=\{\mathcal{X}_H^{(i)}, \mathcal{Y}_H^{(i)}\}_{i=1}^{\mathcal{N}_H}$ . The data augmented based on $\mathcal{D}_H$ is denoted as $\mathcal{D}_{H-\rm{Aug}}$. In the inference stage, users will generate real-time behaviors on the devices. These real-time behavioral data are represented as $\mathcal{D}_R=\{\mathcal{X}_R^{(i)}, \mathcal{Y}_R^{(i)}\}_{i=1}^{\mathcal{N}_R}$. The lagged behavioral data are represented as $\mathcal{D}_{R}^{'}$. $\mathcal{D}_{R}^{'}$ is also data generated by users during the inference stage, and it represents the last set of data uploaded to the cloud by the user. Since the device-to-cloud synchronization is not real-time, there is a time difference between $\mathcal{D}_{R}^{'}$ and $\mathcal{D}_R$. More precisely, $\mathcal{D}_{R}^{'}$ lags behind $\mathcal{D}_R$ by $\Delta t$ time units.

\subsubsection{Model.} LLM and SRM are denoted as $\mathcal{M}_L$ and $\mathcal{M}_S$, respectively. The output of LLM, that is, the candidate list recalled by LLM based on $\mathcal{D}_H$ or $\mathcal{D}_R$, is denoted as $\mathcal{S}=\{v_i\}_{1}^{\mathcal{N}_{\rm{candidate}}}$. The order of items in this list is considered to be the initial ranking, denoted as $\hat{Y}_{\rm{init}}$. The output of SRM re-ranking based on real-time data is denoted as $\hat{Y}_{\rm{rerank}}$.

\subsubsection{Formula.} 
To facilitate the description of the training process, here we use $\phi(\mathcal{D})$ to represent that the model is trained using data $\mathcal{D}$, $\phi_{Re}(\mathcal{D})$ to represent that the model is re-trained using data $\mathcal{D}$, $Aug(\mathcal{D})$ to represent data augmentation based on data $\mathcal{D}$, $\bigoplus$ to represent collaborative algorithms, $\Pi$ to represent the function that determines the inconsistency or similarity of the sorting of two lists. We formulate our proposed LSC4Rec as follows, 
\end{sloppypar}

\textbf{Collaborative Training}:
\begin{equation}
\begin{split}
& \hspace{3em} \mathcal{M}_L=\phi(\mathcal{D}_H),\mathcal{M}_S=\phi(\mathcal{D}_H);\\
& \mathcal{M}_S=\phi_{Re}(\{\mathcal{D}_H,\mathcal{D}_{H-\rm{Aug}}=Aug(\mathcal{M}_L(\mathcal{D}_H))\});\\
& \hspace{7em} \mathcal{M}_S=\phi(\mathcal{D}_R).
\end{split}
\end{equation}

\textbf{Collaborative Inference}:
\begin{equation}
\begin{split}
& \mathcal{S}, \hat{Y}_{\rm{init}}=\mathcal{M}_L(\mathcal{D}_{R}^{'});\\
& \hspace{0.5em} \hat{Y}_{\rm{rerank}}=\mathcal{M}_S(\mathcal{S};\mathcal{D}_{R});\\
& \hspace{1em} \hat{Y}=\hat{Y}_{\rm{init}}\bigoplus\hat{Y}_{\rm{rerank}}.
\end{split}
\end{equation}

\textbf{Collaborative-Decision Request}:
\begin{equation}
\begin{aligned}
\rm{Request}=\mathds{1}(\Pi(\hat{Y}_{\rm{init}}, \hat{Y}_{\rm{rerank}}) \geq Threshold).
\label{eq:update_mrs}
\end{aligned}
\end{equation}
Note that the function symbols here are rough and are only used to formulate the task description. The details of the formula will be elaborated in subsequent subsections of this section.


\subsection{LSC4Rec}
This section describes our proposed LSC4Rec. As shown in Figure~\ref{fig:method}, our LSC4Rec framework includes (a) Collaborative Training, (b) Collaborative Inference, and (c) Collaborative-Decision Request. Collaborative Training aims to enable SRM to provide benefits to the LSC4Rec framework. Collaborative Inference aims to complement the strengths of LLM and SRM during the inference stage, resulting in more accurate recommendation outcomes. Collaborative-Decision Request targets to determine whether there has been a shift in user preferences, in order to decide whether the candidate list needs to be updated. 

\subsubsection{Retropy of Existing Sequential Recommendation Diagrams}
Here we review the paradigm of sequential recommendation. 

In the training stage, if we disregard the process of data textualization required for LLM4Rec, the method by which LLM4Rec and SRM4Rec obtain loss is the same, and can be formalized as follows,
\begin{equation}
    \mathcal{L}=\sum_{u, v, s, y\in\mathcal{D}_H}l(y, \hat{y}=\mathcal{M}(u, v, s)).
\label{eq:loss_general}
\end{equation}
In the above equation, $l$ represents loss function, $\mathcal{M}$ generically represents both LLM and SRM. $\mathcal{M}(\cdot)$ denotes the output of the model when $\cdot$ is given as input. During gradient backpropagation, LLM4Rec and SRM4Rec differ slightly. LLM4Rec has the option to update the instruction, prompt, and model, while SRM4Rec typically focuses on updating the model itself. 

In the inference stage, if we disregard the process of data textualization required for LLM4Rec, the inference processes of both are essentially consistent.

\subsubsection{Collaborative training.}
Collaborative Training consists of Independent training, Cooperative training, and Adaptive re-training.

\noindent \texttt{Independent training.}
When LLM and SRM are trained separately, the calculation of loss is the same as equation~\ref{eq:loss_general}.
The backpropagation for SRM can be formalized as $\argminA_{\theta_S} L$. In the above formula, $\theta_S$ represents the parameter of $\mathcal{M}_S$. Based on prior knowledge, we fix the instruction and prompt of the LLM and then fine-tune the model's parameters. The backpropagation of LLM can be formalized as $\argminA_{\theta_L} L$. In the above equation, $\theta_L$ represents the parameter of $\mathcal{M}_L$.

\noindent \texttt{Cooperative training.}
After independent training, we input $\mathcal{X}_H$ into the LLM to obtain inference results. For each sample $x$, we infer $\mathcal{N}_{candidate}$ items, forming the candidate list $S$. Furthermore, we freely combine items that are difficult for the LLM to distinguish, that is, items with similar tokens, to obtain an enhanced candidate list $S_{\rm{Aug}}$. Subsequently, we also input each sample $x$ into the SRM, allowing the SRM to re-rank within the candidate list $S_{\rm{Aug}}$. The resulting prediction is denoted as $\hat{y}$. So the loss function can be formulated as,
\begin{equation}
    S = \mathcal{M}_L(x), \quad
    S_{\rm{Aug}} = \text{Augment}(S),
\label{eq:candidate_cotraining}
\end{equation}
\begin{equation}
    \mathcal{L}=\sum\nolimits_{u, v, s, y\in\mathcal{D}_H}l(y, \hat{y}=\mathcal{M}_S(u, v, s)|S_{\rm{Aug}}).
\label{eq:loss_cotraining}
\end{equation}
We fix the model parameters of LLM so that the gradient generated by loss only backpropagates to SRM, so the optimization function is $\argminA_{\theta_S} L$.

\noindent \texttt{Adaptive re-training.}
After LLM and SRM are trained, LLM is deployed in the cloud and SRM on the device. However, as mentioned in the Introduction section, due to network bandwidth limitations and delays in data processing on the device side, it is difficult for LLM to access real-time data. Even if it can access such data, its large number of parameters makes it challenging to train and deploy the model quickly. Therefore, SRM needs to be trained based on real-time data on the device, with the optimization goal being $\argminA_{\theta_S} L$, where the loss function is as follows,
\begin{equation}
    \mathcal{L}=\sum\nolimits_{u, v, s, y\in\mathcal{D}_R}l(y, \hat{y}=\mathcal{M}_S(u, v, s)).
\label{eq:loss_srm_realtime_train}
\end{equation}

\subsubsection{Collaborative inference}
Collaborative Inference consists of Cooperative Inference and Result Fusion.

\noindent \texttt{Cooperative Inference.} After training is completed, LLM and SRM are deployed separately in the cloud and on the device. Here, we introduce how to merge the initial ranking results $\hat{Y}_{init}$ output by LLM with the re-ranking results $\hat{Y}_{\rm{rerank}}$ of SRM during the inference stage.
\begin{equation}
\begin{split}
\mathcal{S}, \hat{Y}_{\rm{init}} & =\mathcal{M}_L(\mathcal{D}_{R}^{'});\\
\hspace{0.5em} \hat{Y}_{\rm{rerank}} & =\mathcal{M}_S(\mathcal{S};\mathcal{D}_{R}).
\end{split}
\end{equation}
\noindent \texttt{Result Fusion.} LLM and SRM have different ranking principles; LLM is generative while SRM is non-generative, leading to different ranking scales for each. Therefore, an important step is normalization, which places $\hat{Y}_{\rm{init}}$ and $\hat{Y}_{\rm{rerank}}$ on the same scale to better merge the ranking results. We use $\hat{P}_{\rm{init}}$ to represent the interaction probability between users and products, and $\hat{Y}_{\rm{init}}$ is also obtained from $\hat{P}_{\rm{init}}$. Similarly, $\hat{P}_{\rm{rerank}}$ is also used to obtain $\hat{Y}_{\rm{rerank}}$. The process of normalization can be formalized as follows,
\begin{equation}
\centering
\left\{
\begin{aligned}
& \hat{P}_{\rm{init}}^{\rm{norm}} = \frac{\hat{P}_{\rm{init}} - \min(\hat{P}_{\rm{init}})}{\max(\hat{P}_{\rm{init}}) - \min(\hat{P}_{\rm{init}})}, \\
& \hat{P}_{\rm{rerank}}^{\rm{norm}} = \frac{\hat{P}_{\rm{rerank}} - \min(\hat{P}_{\rm{rerank}})}{\max(\hat{P}_{\rm{rerank}}) - \min(\hat{P}_{\rm{rerank}})}.
\end{aligned}
\right.
\label{eq:norm}
\end{equation}

After normalization and filtering, we adjust the hyperparameter \(\alpha\) and \(\beta\) to merge $\hat{Y}_{\rm{init}}$ and $\hat{Y}_{\rm{rerank}}$,

\begin{equation}
\centering
\left\{
\begin{aligned}
& \hat{P} = \alpha \cdot \hat{P}_{\rm{init}}^{norm} + (1 - \alpha) \cdot \hat{P}_{\rm{rerank}}^{norm},\\
& \hat{Y} = \text{sort}(S,\hat{P}).
\end{aligned}
\right.
\end{equation}

\begin{table*}[ht]
\caption{Performance comparison of baselines and LSC4Rec. 
}
    \label{tab:main_table}
    \vspace{-0.35cm}
    \centering
    \renewcommand{\arraystretch}{1.05}
    \resizebox{0.9\textwidth}{!}{
    \begin{tabular}{c|c|c|c|c|c|c|c|c|c|c}
    \toprule[2pt]
    \multirow{2}{*}{\textbf{Dataset}} & \multirow{2}{*}{\textbf{Model}} & \multicolumn{9}{c}{\textbf{Metric}} \\ \cline{3-11}
     &  & {NDCG@5}  & {NDCG@10}  & {NDCG@20} & {HR@5}  & 
    {HR@10}  & {HR@20}  & {Precision@5} & {Precision@10} & {Precision@20}  \\
    \midrule
    \midrule
    \multirow{10}{*}{\texttt{Beauty}} & DIN (RT) & 0.0079 & 0.0106 & 0.0135 & 0.0131 & 0.0226 & 0.0311 & 0.0026 & 0.0022 & 0.0018 \\
    & GRU4Rec (RT) & 0.0081 & 0.0105 & 0.0129 & 0.0136 & 0.0212 & 0.0308 & 0.0026 & 0.0022 & 0.0017 \\
    & SASRec (RT) & 0.0066 & 0.0096 & 0.0127 & 0.0111 & 0.0207 & 0.0310 & 0.0022 & 0.0021 & 0.0018 \\
    & P5 (RT) & 0.0227 & 0.0257 & 0.0289 & 0.0305 & 0.0400 & 0.0525 & 0.0061 & 0.0040 & 0.0026 \\
    \cline{2-11}
    & DIN (NRT) 
    & 0.0017 & 0.0026 & 0.0042 & 0.0021 & 0.0039 & 0.0083 & 0.0014 & 0.0014 & 0.0015 \\
    & GRU4Rec (NRT) 
    & 0.0017 & 0.0023 & 0.0031 & 0.0020 & 0.0033 & 0.0054 & 0.0014 & 0.0012 & 0.0010 \\
    & SASRec (NRT) 
    & 0.0014 & 0.0021 & 0.0032 & 0.0018 & 0.0033 & 0.0060 & 0.0013 & 0.0011 & 0.0011 \\
    & P5 (NRT)
    & {0.0087} & {0.0108} & {0.0136} & {0.0132} & {0.0200} & {0.0310} & {0.0027} & {0.0020} & {0.0016} \\
    \cline{2-11}
    & \cellcolor{blue!3}Ours (P5+SASRec)
    & \cellcolor{blue!3}\textbf{0.0094} & \cellcolor{blue!3}\textbf{0.0126} & \cellcolor{blue!3}\textbf{0.0154} & \cellcolor{blue!3}\textbf{0.0150} & \cellcolor{blue!3}\textbf{0.0248} & \cellcolor{blue!3}\textbf{0.0361} & \cellcolor{blue!3}\textbf{0.0030} & \cellcolor{blue!3}\textbf{0.0025} &	\cellcolor{blue!3}\textbf{0.0018} \\
    & \cellcolor{blue!3}Improve
    & \cellcolor{blue!3}8.41\% & \cellcolor{blue!3}16.16\%	& \cellcolor{blue!3}13.45\%	& \cellcolor{blue!3}13.14\%	& \cellcolor{blue!3}23.91\%	& \cellcolor{blue!3}16.59\%	& \cellcolor{blue!3}13.21\%	& \cellcolor{blue!3}24.00\%	& \cellcolor{blue!3}16.77\% \\
    \hline
    \hline
    \multirow{10}{*}{\texttt{Toys}} & DIN (RT) & 0.0046 & 0.0063 & 0.0081 & 0.0076 & 0.0128 & 0.0202 & 0.0015 & 0.0013 & 0.0010 \\
    & GRU4Rec (RT) & 0.0050 & 0.0073 & 0.0098 & 0.0085 & 0.0156 & 0.0253 & 0.0017 & 0.0016 & 0.0013 \\
    & SASRec (RT) & 0.0052 & 0.0073 & 0.0101 & 0.0080 & 0.0148 & 0.0259 & 0.0016 & 0.0015 & 0.0013 \\
    & P5 (RT) & 0.0187 & 0.0204 & 0.0221 & 0.0237 & 0.0291 & 0.0358 & 0.0047 & 0.0029 & 0.0018 \\
    \cline{2-11}
    & DIN (NRT) 
    & 0.0007 & 0.0010 & 0.0017 & 0.0010 & 0.0016 & 0.0035 & 0.0006 & 0.0005 & 0.0006 \\
    & GRU4Rec (NRT) 
    & 0.0009 & 0.0015 & 0.0020 & 0.0011 & 0.0023 & 0.0037 & 0.0008 & 0.0008 & 0.0007 \\
    & SASRec (NRT) 
    & 0.0015 & 0.0020 & 0.0026 & 0.0019 & 0.0030 & 0.0046 & 0.0012 & 0.0010 & 0.0008 \\
    & P5 (NRT) & {0.0066}	& {0.0078}	& {0.0090}	& {0.0096}	& {0.0133}	& {0.0183}	& {0.0019}	& {0.0013}	& {0.0009} \\
    \cline{2-11}
    & \cellcolor{blue!3}Ours (P5+SASRec)
    & \cellcolor{blue!3}\textbf{0.0066}	& \cellcolor{blue!3}\textbf{0.0084}	& \cellcolor{blue!3}\textbf{0.0096}	& \cellcolor{blue!3}\textbf{0.0105}	& \cellcolor{blue!3}\textbf{0.0158}	& \cellcolor{blue!3}\textbf{0.0209}	& \cellcolor{blue!3}\textbf{0.0021}	& \cellcolor{blue!3}\textbf{0.0016}	& \cellcolor{blue!3}\textbf{0.0010} \\
    & \cellcolor{blue!3}Improve
    & \cellcolor{blue!3}0.46\%	& \cellcolor{blue!3}7.33\%	& \cellcolor{blue!3}6.76\%	& \cellcolor{blue!3}8.62\%	& \cellcolor{blue!3}18.52\%	& \cellcolor{blue!3}13.74\%	& \cellcolor{blue!3}8.29\%	& \cellcolor{blue!3}18.80\%	& \cellcolor{blue!3}13.04\% \\
    \hline
    \hline
    \multirow{10}{*}{\texttt{Yelp}} 
    & DIN (RT) & 0.0067 & 0.0094 & 0.0126 & 0.0108 & 0.0191 & 0.0318 & 0.0022 & 0.0019 & 0.0016 \\
    & GRU4Rec (RT) & 0.0045 & 0.0063 & 0.0084 & 0.0075 & 0.0131 & 0.0217 & 0.0015 & 0.0013 & 0.0011 \\
    & SASRec (RT) & 0.0056 & 0.0079 & 0.0110 & 0.0087 & 0.0159 & 0.0281 & 0.0017 & 0.0016 & 0.0014 \\
    & P5 (RT) & 0.0226 & 0.0253 & 0.0276 & 0.0299 & 0.0382 & 0.0471 & 0.0060 & 0.0038 & 0.0024 \\
    \cline{2-11}
    & DIN (NRT) 
    &
    0.0033 & 0.0044 & 0.0057 & 0.0040 & 0.0062 & 0.0096 & 0.0026 & \underline{0.0020} & \underline{0.0014} \\
    & GRU4Rec (NRT) 
    &
    0.0017 & 0.0026 & 0.0036 & 0.0022 & 0.0041 & 0.0065 & 0.0013 & 0.0013 & 0.0011 \\
    & SASRec (NRT) 
    &
    0.0021 & 0.0030 & 0.0039 & 0.0024 & 0.0044 & 0.0066 & 0.0016 & 0.0015 & 0.0012 \\
    & P5 (NRT)
    & {0.0093}	& {0.0112}	& {0.0132}	& {0.0136}	& {0.0196}	& {0.0272}	& {0.0027}	& {0.0020}	& {0.0014} \\
    \cline{2-11}
    & \cellcolor{blue!3}Ours (P5+SASRec)
    & \cellcolor{blue!3}\textbf{0.0095}	& \cellcolor{blue!3}\textbf{0.0116}	& \cellcolor{blue!3}\textbf{0.0144}	& \cellcolor{blue!3}\textbf{0.0145}	& \cellcolor{blue!3}\textbf{0.0213}	& \cellcolor{blue!3}\textbf{0.0323}	& \cellcolor{blue!3}\textbf{0.0029}	& \cellcolor{blue!3}\textbf{0.0021}	& \cellcolor{blue!3}\textbf{0.0016} \\
    & \cellcolor{blue!3}Improve
    & \cellcolor{blue!3}1.72\%	& \cellcolor{blue!3}3.65\%	& \cellcolor{blue!3}9.50\%	& \cellcolor{blue!3}7.00\%	& \cellcolor{blue!3}8.88\%	& \cellcolor{blue!3}18.85\%	& \cellcolor{blue!3}7.01\%	& \cellcolor{blue!3}8.67\%	& \cellcolor{blue!3}19.12\% \\
    \bottomrule[2pt]
    \end{tabular}
    }
    \vspace{-0.2cm}
\end{table*}
\subsubsection{Collaborative-Decision Request}
After collaborative training and collaborative inference, the LSC4Rec has formed a complete framework. This framework can compensate for the performance decline due to LLM's difficulty in obtaining real-time data as much as possible through SRM, making LSC4Rec more effective than using LLM4Rec or SRM4Rec independently in this system. However, a noteworthy point is that when LSC4Rec should invoke LLM to provide new inference results based on real-time data has not been addressed. To fill this gap, we designed the Collaborative-Decision Request feature for LSC4Rec.

We utilize the positional difference in rankings of the same element by $\hat{y}_{\rm{init}}$ and $\hat{y}_{\rm{rerank}}$ to calculate a inconsistency score $c$. Assuming $\rm{pos}_p(q)$ represents the position of element $q$ in list $p$, the inconsistency score $c$ can be formalized as follows,
\begin{equation}
    c = \frac{1}{n} \sum\nolimits_{u \in \hat{y}_{\rm{init}} \cap \hat{y}_{\rm{rerank}}} |\rm{pos}_{\hat{y}_{\rm{init}}}(u) - \rm{pos}_{\hat{y}_{\rm{rerank}}}(u)|,
\label{eq:inconsistency}
\end{equation}
where $\hat{y}_{\rm{init}}$ and $\hat{y}_{\rm{rerank}}$
In the above formula, $\hat{y}_{\rm{init}} \cap \hat{y}_{\rm{rerank}} = \hat{y}_{\rm{init}} = \hat{y}_{\rm{rerank}}$. This is because the elements in lists $\hat{y}_{\rm{init}}$ and $\hat{y}_{\rm{rerank}}$ are exactly the same, only their order differs.
\begin{equation}
\begin{aligned}
   \rm{Request}=\mathds{1}(c \geq Threshold).
\label{eq:request}
\end{aligned}
\end{equation}
In the equation above, $\mathds{1}(\cdot)$ is the indicator function. 
To get the threshold, we need to collect user data for a period of time, then get the inconsistency $c$ corresponding to these data on the cloud and sort them, and then set the threshold according to the load of the cloud server. That is, the threshold can be obtained during the training on the training set. For example, if the load of the cloud server needs to be reduced by 90\%, that is, when the load is only 10\% of the previous value, only the minimum 10\% position value needs to be sent to each device as the threshold. During the inference process, each device determines whether it needs to upload real-time data to the LLM for inference, based on formulas \ref{eq:inconsistency} and \ref{eq:request}. This is done to ensure the most optimal LLM invocation under any device-cloud communication resources and LLM invocation resources.
\section{Experiments}
\label{sec:experiments}

\renewcommand{\algorithmicrequire}{\textbf{Input:}}
\renewcommand{\algorithmicrequire}{\textbf{Initialization:}}
\begin{table*}[!ht]
    \centering
    \caption{The impact of choosing different LLMs on performance.}
    \label{tab:analysis_different_llm}
    \vspace{-0.3cm}
    \renewcommand{\arraystretch}{0.98}
    \resizebox{0.86\textwidth}{!}{
    \begin{tabular}{c|c|c|c|c|c|c|c|c|c|c}
    \toprule[2pt]
    \multirow{2}{*}{\textbf{Dataset}} & \multirow{2}{*}{\textbf{LLM}} & \multicolumn{9}{c}{\textbf{Metric}} \\ \cline{3-11}
     &  & {NDCG@5}  & {NDCG@10}  & {NDCG@20} & {HR@5}  & 
    {HR@10}  & {HR@20}  & {Precision@5} & {Precision@10} & {Precision@20}  \\
    \midrule
    \midrule
    \multirow{3}{*}{\texttt{Beauty}} & POD (NRT) & 0.0079 & 0.0111 & 0.0134 & 0.0139 & 0.0239 & 0.0327 & 0.0028 & 0.0024 & 0.0016 \\
    & Ours(POD+SASRec) & \textbf{0.0091} & \textbf{0.0125} & \textbf{0.0150} & \textbf{0.0150} & \textbf{0.0256} & \textbf{0.0356} & \textbf{0.0030} & \textbf{0.0026} & \textbf{0.0018} \\
    & Improv & 15.59\% & 12.41\% & 12.27\% & 8.37\% & 7.29\% & 8.62\% & 8.30\% & 7.11\% & 8.54\% \\
    \bottomrule[2pt]
    \end{tabular}
    }
    \vspace{-0.2cm}
\end{table*}
\renewcommand{\algorithmicrequire}{\textbf{Input:}}
\renewcommand{\algorithmicrequire}{\textbf{Initialization:}}
\begin{table*}[!ht]
    \centering
    \caption{The impact of choosing different SRMs on performance.}
    \label{tab:analysis_different_srm}
    \vspace{-0.35cm}
    \renewcommand{\arraystretch}{0.95}
    \resizebox{0.82\textwidth}{!}{
    \begin{tabular}{c|c|c|c|c|c|c|c|c|c|c}
    \toprule[2pt]
    \multirow{2}{*}{\textbf{Dataset}} & \multirow{2}{*}{\textbf{SRM}} & \multicolumn{9}{c}{\textbf{Metric}} \\ \cline{3-11}
     &  & {NDCG@5}  & {NDCG@10}  & {NDCG@20} & {HR@5}  & 
    {HR@10}  & {HR@20}  & {Precision@5} & {Precision@10} & {Precision@20}  \\
   
    \midrule
    \midrule
    
    \multirow{3}{*}{\texttt{Beauty}} & {DIN} 
    & 0.0091 & 0.0118 & 0.0147 & 0.0145 & 0.0230 & 0.0343 & 0.0029 & 0.0023 & 0.0017 \\
    & {GRU4Rec} 
    & 0.0092 & 0.0124 & 0.0151 & 0.0146 & 0.0247 & 0.0353 & 0.0029 & 0.0024 & 0.0017 \\
    & {SASRec} 
    & \textbf{0.0094} & \textbf{0.0126} & \textbf{0.0154} & \textbf{0.0150} & \textbf{0.0248} & \textbf{0.0361} & \textbf{0.0030} & \textbf{0.0025} & \textbf{0.0018} \\
    \bottomrule[2pt]
    \end{tabular}
    }
    \vspace{-0.2cm}
\end{table*}

\begin{table*}[!ht]
\caption{
The ablation studies of the collaborative training.}
    \label{tab:ablation_training}
    \vspace{-0.3cm}
    \centering
    \renewcommand{\arraystretch}{0.9}
    \resizebox{0.87\textwidth}{!}{
    \begin{tabular}{c|c|c|c|c|c|c|c|c|c|c|c}
    \toprule[2pt]
     & \multicolumn{2}{c|}{\textbf{Training Method}} & \multicolumn{9}{c}{\textbf{Metric}} \\ \cline{2-12}
    \multirow{-2}{*}{\textbf{Dataset}} & {\makecell{Collaborative \\ Training}}  & {\makecell{Adaptive \\ Re-training}}  & {NDCG@5}  & {NDCG@10}  & {NDCG@20} & {HR@5}  & 
    {HR@10}  & {HR@20}  & {Precision@5} & {Precision@10} & {Precision@20}  \\
        \midrule
       \midrule
    \multirow{4}{*}{\texttt{Beauty}} 
    & \XSolidBrush & \XSolidBrush
    & 0.0073 & 0.0101 & 0.0130 & 0.0116 & 0.0204 & 0.0322 & 0.0023 & 0.0020 & 0.0016  \\
    & \Checkmark & \XSolidBrush
    & 0.0089 & 0.0113 & 0.0138 & 0.0143 & 0.0216 & 0.0318 & \underline{0.0029} & 0.0022 & 0.0016 \\
    & \XSolidBrush & \Checkmark
    & \underline{0.0093} & \underline{0.0122} & \underline{0.0150} & \underline{0.0145} & \underline{0.0236} & \underline{0.0347} & \underline{0.0029} & \underline{0.0024} & \underline{0.0017} \\
    & \cellcolor{blue!3}\Checkmark & \cellcolor{blue!3}\Checkmark
    & \cellcolor{blue!3}\textbf{0.0094} & \cellcolor{blue!3}\textbf{0.0126} & \cellcolor{blue!3}\textbf{0.0154} & \cellcolor{blue!3}\textbf{0.0150} & \cellcolor{blue!3}\textbf{0.0248} & \cellcolor{blue!3}\textbf{0.0361} & \cellcolor{blue!3}\textbf{0.0030} & \cellcolor{blue!3}\textbf{0.0025} & \cellcolor{blue!3}\textbf{0.0018} \\
       \midrule
    \multirow{4}{*}{\texttt{Toys}}
    & \XSolidBrush & \XSolidBrush
    & 0.0047 & 0.0065 & 0.0082 & 0.0074 & 0.0130 & 0.0194 & 0.0015 & 0.0013 & 0.0009 \\
    & \Checkmark & \XSolidBrush
    & 0.0062 & 0.0074 & 0.0087 & 0.0098 & 0.0136 & 0.0189 & 0.0020 & 0.0014 & 0.0009 \\
    & \XSolidBrush & \Checkmark
    & \underline{0.0064} & \underline{0.0080} & \underline{0.0095} & \underline{0.0099} & \underline{0.0146} & \underline{0.0208} & \underline{0.0020} & \underline{0.0015} & \textbf{0.0010} \\
    & \cellcolor{blue!3}\Checkmark & \cellcolor{blue!3}\Checkmark
    & \cellcolor{blue!3}\textbf{0.0066} & \cellcolor{blue!3}\textbf{0.0084} & \cellcolor{blue!3}\textbf{0.0096} & \cellcolor{blue!3}\textbf{0.0105} & \cellcolor{blue!3}\textbf{0.0158} & \cellcolor{blue!3}\textbf{0.0209} & \cellcolor{blue!3}\textbf{0.0021} & \cellcolor{blue!3}\textbf{0.0016} & \cellcolor{blue!3}\textbf{0.0010} \\
    \midrule 
    \multirow{4}{*}{\texttt{Yelp}}
    & \XSolidBrush & \XSolidBrush
    & \textbf{0.0068} & 0.0090 & 0.0120 & \textbf{0.0104} & 0.0173 & 0.0291 & 0.0021 & 0.0017 & 0.0014 \\
    & \Checkmark & \XSolidBrush
    & {\underline{0.0089}} & {\underline{0.0110}} & 0.0130 & {\underline{0.0137}} & 0.0202 & 0.0279 & {\underline{0.0027}} & 0.0019 & 0.0014 \\
    & \XSolidBrush & \Checkmark
    & 0.0083 & 0.0109 & {\underline{0.0136}} & 0.0131 & {\underline{0.0209}} & {\underline{0.0317}} & 0.0026 & {\underline{0.0020}} & {\underline{0.0015}} \\
    & \cellcolor{blue!3}\Checkmark & \cellcolor{blue!3}\Checkmark
    & \cellcolor{blue!3}\textbf{0.0095} & \cellcolor{blue!3}\textbf{0.0116} & \cellcolor{blue!3}\textbf{0.0144} & \cellcolor{blue!3}\textbf{0.0145} & \cellcolor{blue!3}\textbf{0.0213} & \cellcolor{blue!3}\textbf{0.0323} & \cellcolor{blue!3}\textbf{0.0029} & \cellcolor{blue!3}\textbf{0.0021} & \cellcolor{blue!3}\textbf{0.0016} \\
    \bottomrule[2pt]
    \end{tabular}
    }
    \vspace{-0.2cm}
\end{table*}

\subsection{Experimental Setup}

\subsubsection{Datasets.}
We evaluated our method on three widely used datasets, which
are all collected from an e-commerce platform Amazon\footnote{\url{https://jmcauley.ucsd.edu/data/amazon/}\label{fn:amazon}}  and cover various product categories such as books, electronics, home goods, and more, they usually include user reviews, ratings, product descriptions, prices, and other information. We also evaluated our method on Yelp\footnote{\url{https://www.yelp.com/dataset}\label{fn:yelp}} which contains details about businesses, user reviews, ratings, business categories, addresses, and more. In accordance with convention, all user-item pairs in the datasets were considered as positive samples. For the training and test sets, user-item pairs that did not exist in the datasets were sampled as negative samples~\cite{ref:llm4rec_P5,ref:llm4rec_Prompt_Distillation}. During testing, we rank on a sample set consisting of all items. In these datasets, we use a user's last behavior for testing, the second to last behavior for validation, and the remaining behaviors for training.

\subsubsection{Baselines.}
\label{subsec:experiment_baseline}
We used the following SRMs and LLMs for the experiments.
\begin{itemize}[itemsep=1pt,topsep=2pt,leftmargin=20pt]
    \item \textbf{SRMs}. \textit{DIN}~\cite{ref:din}, \textit{GRU4Rec}~\cite{ref:gru4rec}, and \textit{SASRec}~\cite{ref:sasrec} are three highly prevalent SRMs in both academic research and the industry. They each incorporate different techniques, such as GRU (Gated Recurrent Unit), Attention, and Self-Attention, to enhance the recommendation process. One point to note is that although models like BERT4Rec and E4SRec achieve better performance, their significantly larger number of model parameters results in far greater resource consumption compared to the aforementioned lightweight SRMs. Therefore, we choose not to use them as models on the device.\\
    \begin{sloppypar}
    \item \textbf{LLMs}. \textit{P5}~\cite{ref:llm4rec_P5} and \textit{POD}~\cite{ref:llm4rec_Prompt_Distillation} are two well-known LLM paradigms for recommendation, aiming to transform the recommendation task into textual format and use the text as input to the LLM to obtain recommendation results.  \\ 
    \end{sloppypar}
\end{itemize}
\textit{In the subsequent experimental results, unless otherwise specified, LLM defaults to P5 and SRM defaults to SASRec.}

\begin{table*}[!ht]
\caption{
The ablation studies of the collaborative inference.}
    \label{tab:ablation_inference}
    \vspace{-0.3cm}
    \centering
 \renewcommand{\arraystretch}{0.9}
    \resizebox{0.85\textwidth}{!}{
    \begin{tabular}{c|c|c|c|c|c|c|c|c|c|c|c|c}
    \toprule[2pt]
    \multirow{3}{*}{\textbf{Dataset}} & \multicolumn{3}{c|}{\textbf{Inference Method}} & \multicolumn{9}{c}{\textbf{Metric}} \\ \cline{2-13}
     & {LLM}  & {SRM} & {\makecell{\makecell{Result \\ Fusion}}}  & {NDCG@5}  & {NDCG@10}  & {NDCG@20} & {HR@5}  & 
    {HR@10}  & {HR@20}  & {Precision@5} & {Precision@10} & {Precision@20}  \\
    \midrule
   \midrule
    \multirow{4}{*}{\texttt{Beauty}} 
    & \Checkmark & \XSolidBrush & \XSolidBrush
    & 0.0087 & 0.0108 & 0.0136 & 0.0132 & 0.0200 & 0.0310 & 0.0027 & 0.0020 & \underline{0.0016} \\
    & \XSolidBrush & \Checkmark & \XSolidBrush
    & 0.0014 & 0.0021 & 0.0032 & 0.0018 & 0.0033 & 0.0060 & 0.0013 & 0.0011 & 0.0011 \\
    & \Checkmark & \Checkmark & \XSolidBrush
    & \textbf{0.0094} & \underline{0.0122} & \underline{0.0144} & \textbf{0.0150} & \underline{0.0238} & \underline{0.0329} & \underline{0.0030} & \underline{0.0024} & \underline{0.0016} \\
    & \cellcolor{blue!3}\Checkmark & \cellcolor{blue!3}\Checkmark & \cellcolor{blue!3}\Checkmark
    & \cellcolor{blue!3}\textbf{0.0094} & \cellcolor{blue!3}\textbf{0.0126} & \cellcolor{blue!3}\textbf{0.0154} & \cellcolor{blue!3}\textbf{0.0150} & \cellcolor{blue!3}\textbf{0.0248} & \cellcolor{blue!3}\textbf{0.0361} & \cellcolor{blue!3}\textbf{0.0030} & \cellcolor{blue!3}\textbf{0.0025} & \cellcolor{blue!3}\textbf{0.0018} \\
       \midrule
    \multirow{4}{*}{\texttt{Toys}}
    & \Checkmark & \XSolidBrush & \XSolidBrush
    & \underline{0.0087} & \underline{0.0108} & \underline{0.0136} & \underline{0.0132} & \underline{0.0200} & \underline{0.0310} & \underline{0.0027} & \underline{0.0020} & \underline{0.0016} \\
    & \XSolidBrush & \Checkmark & \XSolidBrush
    & 0.0015 & 0.0020 & 0.0026 & 0.0019 & 0.0030 & 0.0046 & 0.0012 & 0.0010 & 0.0008 \\
    & \Checkmark & \Checkmark & \XSolidBrush
    & 0.0059 & 0.0074 & 0.0089 & 0.0092 & 0.0138 & 0.0198 & 0.0018 & 0.0014 & 0.0010 \\
    & \cellcolor{blue!3}\Checkmark & \cellcolor{blue!3}\Checkmark & \cellcolor{blue!3}\Checkmark
    & \cellcolor{blue!3}\textbf{0.0094} & \cellcolor{blue!3}\textbf{0.0126} & \cellcolor{blue!3}\textbf{0.0154} & \cellcolor{blue!3}\textbf{0.0150} & \cellcolor{blue!3}\textbf{0.0248} & \cellcolor{blue!3}\textbf{0.0361} & \cellcolor{blue!3}\textbf{0.0030} & \cellcolor{blue!3}\textbf{0.0025} & \cellcolor{blue!3}\textbf{0.0018} \\
    \midrule
    \multirow{4}{*}{\texttt{Yelp}} 
    & \Checkmark & \XSolidBrush & \XSolidBrush
    & \underline{0.0087} & \underline{0.0108} & \underline{0.0136} & \underline{0.0132} & 0.0200 & 0.0310 & \underline{0.0027} & 0.0020 & \underline{0.0016} \\
    & \XSolidBrush & \Checkmark & \XSolidBrush
    & 0.0041 & 0.0058 & 0.0083 & 0.0066 & 0.0120 & 0.0218 & 0.0013 & 0.0012 & 0.0011 \\
    & \Checkmark & \Checkmark & \XSolidBrush
    & 0.0083 & 0.0108 & 0.0134 & 0.0127 & \underline{0.0205} & \underline{0.0311} & 0.0025 & \underline{0.0021} & \underline{0.0016} \\
    & \cellcolor{blue!3}\Checkmark & \cellcolor{blue!3}\Checkmark & \cellcolor{blue!3}\Checkmark
    & \cellcolor{blue!3}\textbf{0.0094} & \cellcolor{blue!3}\textbf{0.0126} & \cellcolor{blue!3}\textbf{0.0154} & \cellcolor{blue!3}\textbf{0.0150} & \cellcolor{blue!3}\textbf{0.0248} & \cellcolor{blue!3}\textbf{0.0361} & \cellcolor{blue!3}\textbf{0.0030} & \cellcolor{blue!3}\textbf{0.0025} & \cellcolor{blue!3}\textbf{0.0018} \\
    \bottomrule[2pt]
    \end{tabular}
    }
    \vspace{-0.3cm}
\end{table*}

\subsubsection{Evaluation metrics} 
In the experiments, we use the widely adopted NDCG, HitRate(HR) and Precision as the metrics to evaluate model performance. The details of the metrics are in Appendix.

\subsection{Experimental Results}
We use bold to denote the \textbf{best} value and underline to denote the \uline{second-best} value (if applicable). Typically, we conduct such comparisons for each dataset. However, in some cases, such as in Figure~\ref{tab:ablation_inference}, we compare values for each Request Frequency. In the experimental analysis section, when results across all datasets would take up significant space and consistent conclusions are drawn for each dataset, we only present the results for one dataset (e.g., Figure~\ref{tab:ablation_inference}) to save space for more critical content.

\subsubsection{Overall performance}
Table~\ref{tab:main_table} presents the performance comparison between LSC4Rec, SRMs, and LLMs under both real-time (RT) and near-real-time (NRT) data settings. Compared to real-time data, near-real-time data for each user excludes the most recent two interactions. As our study focuses on sequential recommendation tasks, we trained and tested LLMs exclusively on sequential recommendation datasets, excluding other types of datasets such as rating prediction.
For LSC4Rec, LLMs first receive the same two-click-delayed near-real-time data to generate a candidate list of length 50. This candidate list is then re-ranked by SRMs using real-time data. In our setup, LLMs are instantiated as P5 or POD, and SRMs are set to SASRec (the evaluation of collaboration with other SRMs can be found in Table~\ref{tab:analysis_different_srm}). From Table~\ref{tab:main_table}, we can draw the following conclusions:

\begin{sloppypar}
\begin{itemize}[itemsep=1pt,topsep=2pt,leftmargin=20pt]
    \item \textbf{Necessity of LLMs:} Under various conditions of accessing real-time data, the performance of LLMs is superior to that of SRMs in most cases, highlighting the necessity of utilizing on-cloud LLMs.
    \item \textbf{Necessity of SRMs:}
    Although the performance of SRM based on real-time data sometimes surpasses that of LSC4Rec in some metrics, achieving this in practical applications is quite challenging. This is because SRMs often cannot access real-time data (if deployed on the cloud) and all of the item embeddings (if deployed on the device). Nevertheless, the high performance of SRM with real-time data highlights the unique advantages that on-device SRMs bring to LSC4Rec, which LLMs lack.
    \item \textbf{Impact of Real-Time Data Absence on LLMs and SRMs:} Compared to real-time data, the performance of both LLMs and SRMs drops significantly when using near-real-time data delayed by two user interactions. This underscores the negative impact of lacking access to real-time user behavior on models, especially on cloud-based LLMs.
    \item \textbf{Effectiveness of LSC4Rec:} Our LSC4Rec method achieves significant performance improvements across all three datasets. Specifically, when LLM is set to P5 and SRM to SASRec, LSC4Rec achieves average improvements of 16.18\%, 10.62\%, and 9.38\% on the Beauty, Toys, and Yelp datasets, respectively. These results indicate that our method effectively mitigates the performance degradation of LLMs caused by the lack of access to real-time user behavior.
\end{itemize}
\end{sloppypar}

\begin{table*}[!ht]
\caption{
The ablation studies of the collaborative-decision request.}
    \label{tab:ablation_request}
    \vspace{-0.3cm}
\resizebox{0.92\textwidth}{!}{
\begin{tabular}{c|c|c|c|c|c|c|c|c|c|c|c}
\toprule[2pt]
 &  &  & \multicolumn{9}{c}{\textbf{Metric}} \\
\cline{4-12}
\multirow{-2}{*}{\textbf{Dataset}} & \multirow{-2}{*}{\textbf{\makecell{Decision\\Method}}} & \multirow{-2}{*}{\textbf{\makecell{Request\\Frequency}}} & NDCG@5 & NDCG@10 & NDCG@20 & HR@5 & HR@10 & HR@20 & Precision@5 & Precision@10 & Precision@20 \\
\midrule
\midrule
 & Random & 5\% & 0.0091 & 0.0114 & 0.0142 & 0.0138 & 0.0209 & 0.0321 & 0.0028 & 0.0021 & 0.0016 \\
 & \cellcolor{blue!3}Inconsistency & \cellcolor{blue!3}5\% & \cellcolor{blue!3}\textbf{0.0125} & \cellcolor{blue!3}\textbf{0.0146} & \cellcolor{blue!3}\textbf{0.0173} & \cellcolor{blue!3}\textbf{0.0173} & \cellcolor{blue!3}\textbf{0.0240} & \cellcolor{blue!3}\textbf{0.0349} & \cellcolor{blue!3}\textbf{0.0035} & \cellcolor{blue!3}\textbf{0.0024} & \cellcolor{blue!3}\textbf{0.0017} \\ 
 & \multicolumn{2}{c|}{\cellcolor{blue!3}Improvement} & \cellcolor{blue!3}36.66\% & \cellcolor{blue!3}28.14\% & \cellcolor{blue!3}21.89\% & \cellcolor{blue!3}25.71\% & \cellcolor{blue!3}14.80\% & \cellcolor{blue!3}8.63\% & \cellcolor{blue!3}25.82\% & \cellcolor{blue!3}14.83\% & \cellcolor{blue!3}8.07\% \\ \cline{2-12}
 & Random & 10\% & 0.0097 & 0.0120 & 0.0149 & 0.0145 & 0.0219 & 0.0332 & 0.0029 & 0.0022 & 0.0017 \\
 & \cellcolor{blue!3}Inconsistency & \cellcolor{blue!3}10\% & \cellcolor{blue!3}\textbf{0.0134} & \cellcolor{blue!3}\textbf{0.0156} & \cellcolor{blue!3}\textbf{0.0183} & \cellcolor{blue!3}\textbf{0.0184} & \cellcolor{blue!3}\textbf{0.0254} & \cellcolor{blue!3}\textbf{0.0363} & \cellcolor{blue!3}\textbf{0.0037} & \cellcolor{blue!3}\textbf{0.0025} & \cellcolor{blue!3}\textbf{0.0018} \\ 
 & \multicolumn{2}{c|}{\cellcolor{blue!3}Improvement} & \cellcolor{blue!3}37.91\% & \cellcolor{blue!3}29.56\% & \cellcolor{blue!3}23.13\% & \cellcolor{blue!3}26.50\% & \cellcolor{blue!3}15.91\% & \cellcolor{blue!3}9.31\% & \cellcolor{blue!3}26.46\% & \cellcolor{blue!3}15.98\% & \cellcolor{blue!3}9.04\% \\ \cline{2-12}
 & Random & 20\% & 0.0104 & 0.0127 & 0.0157 & 0.0153 & 0.0227 & 0.0344 & 0.0031 & 0.0023 & 0.0017 \\
 & \cellcolor{blue!3}Inconsistency & \cellcolor{blue!3}20\% & \cellcolor{blue!3}\textbf{0.0140} & \cellcolor{blue!3}\textbf{0.0162} & \cellcolor{blue!3}\textbf{0.0190} & \cellcolor{blue!3}\textbf{0.0192} & \cellcolor{blue!3}\textbf{0.0263} & \cellcolor{blue!3}\textbf{0.0373} & \cellcolor{blue!3}\textbf{0.0038} & \cellcolor{blue!3}\textbf{0.0026} & \cellcolor{blue!3}\textbf{0.0019} \\ 
 & \multicolumn{2}{c|}{\cellcolor{blue!3}Improvement} & \cellcolor{blue!3}34.72\% & \cellcolor{blue!3}27.42\% & \cellcolor{blue!3}21.19\% & \cellcolor{blue!3}25.44\% & \cellcolor{blue!3}15.54\% & \cellcolor{blue!3}8.32\% & \cellcolor{blue!3}25.49\% & \cellcolor{blue!3}15.42\% & \cellcolor{blue!3}8.14\% \\ \cline{2-12}
 & Random & 40\% & 0.0137 & 0.0162 & 0.0192 & 0.0195 & 0.0274 & 0.0393 & 0.0039 & 0.0027 & 0.0020 \\
 & \cellcolor{blue!3}Inconsistency & \cellcolor{blue!3}40\% & \cellcolor{blue!3}\textbf{0.0164} & \cellcolor{blue!3}\textbf{0.0189} & \cellcolor{blue!3}\textbf{0.0219} & \cellcolor{blue!3}\textbf{0.0225} & \cellcolor{blue!3}\textbf{0.0305} & \cellcolor{blue!3}\textbf{0.0423} & \cellcolor{blue!3}\textbf{0.0045} & \cellcolor{blue!3}\textbf{0.0031} & \cellcolor{blue!3}\textbf{0.0021} \\ 
\multirow{-12}{*}{\texttt{Beauty}} & \multicolumn{2}{c|}{\cellcolor{blue!3}Improvement} & \cellcolor{blue!3}19.82\% & \cellcolor{blue!3}16.92\% & \cellcolor{blue!3}14.18\% & \cellcolor{blue!3}15.89\% & \cellcolor{blue!3}11.58\% & \cellcolor{blue!3}7.74\% & \cellcolor{blue!3}15.94\% & \cellcolor{blue!3}11.31\% & \cellcolor{blue!3}8.16\% \\
\bottomrule[2pt]
\end{tabular}
}
\vspace{-0.2cm}
\end{table*}

\subsubsection{The impact of choosing different LLMs on performance.}
To observe the impact of small model selection on LSC4Rec, we conducted experiments using various small models. As shown in Table~\ref{tab:analysis_different_llm}, the experimental results indicate that: LSC4Rec achieves significant performance improvements on POD, indicating that the LSC4Rec framework can effectively mitigate the performance degradation of various LLMs caused by their inability to access real-time data.

\subsubsection{The impact of choosing different SRMs on performance.}
To observe the impact of small model selection on LSC4Rec, we conducted experiments based on  various small models. As shown in Table~\ref{tab:analysis_different_srm}, the experimental results indicate that:

\begin{itemize}
    \item The choice of small models indeed has some influence on LSC4Rec, which can be attributed to differences in the ability of different SRMs to interpret user preferences. Overall, SASRec provides the most significant improvement for LSC4Rec. We conducted analyses on the Beauty, Toys, and Yelp datasets. Although SASRec does not outperform GRU4Rec and DIN across all metrics on all datasets, it generally provides the most significant improvements to LSC4Rec. Therefore, to save space, we present only the results on the Beauty dataset.
    \item Within the LSC4Rec framework, regardless of the SRM used, the performance is superior to that of LLMs without access to real-time data.
\end{itemize}

\subsection{Ablation Study}
For simplicity, we use ``\textbf{w.}'' to represent ``with'' and ``\textbf{w/o.}'' to represent ``without''.
In these experiments, the SRM on the device uses the SASRec model.

\subsubsection{The impact of collaborative training.}
As shown in Table~\ref{tab:ablation_training}, we analyze the effectiveness of each component in collaborative training via ablation study. There are four rows for each dataset, corresponding to four types of ablations, we provide the following explanations for them:
\begin{sloppypar}
\begin{itemize}[itemsep=1pt,topsep=2pt,leftmargin=20pt]
    \item \textbf{w/o.} Collaborative Training and \textbf{w/o.} Adaptive Re-training (Row.1) indicates that both the LLMs and SRMs are trained independently based on historical user data. 
    \item \textbf{w.} Collaborative Training and \textbf{w/o.} Adaptive Re-training (Row.2) indicates that after both the LLMs and SRMs are trained independently based on historical user data, the SRM then learns to re-rank the candidate list generated by the LLM.
    \item \textbf{w/o.} Collaborative Training and \textbf{w.} Adaptive Re-training (Row.3) indicates that after both the LLMs and SRMs are trained independently based on historical user data, the SRM is then re-trained on the device based on the user's real-time data.
    \item \textbf{w.} Collaborative Training and \textbf{w.} Adaptive Re-training (Row.4) indicates that after both the LLMs and SRMs are trained independently based on historical user data, the SRM learns to re-rank the candidate list generated by the LLM. Then the SRM is re-trained on the device based on the user's real-time data.
\end{itemize}
\end{sloppypar}
The experimental results show that with cooperative training and adaptive re-training performs the best, followed by with adaptive re-training and without cooperative training. The worst case occurs when neither is used, and the gap compared to the other three cases is significant. This indicates that on-device SRM cannot perform well within the LSC4Rec framework solely through independent training. These results demonstrate the effectiveness of cooperative training and adaptive re-training.

\begin{table*}[!ht]
    \centering
    \caption{The impact of length of candidate list on performance}
    \label{tab:analysis_candidate_length}
    \vspace{-0.3cm}
    \renewcommand{\arraystretch}{1.05}
    \resizebox{0.85\textwidth}{!}{
    \begin{tabular}{c|c|c|c|c|c|c|c|c|c|c}
    \toprule[2pt]
    \multirow{2}{*}{\textbf{Dataset}} & \multirow{2}{*}{\textbf{\makecell{Candidate\\Length}}} & \multicolumn{9}{c}{\textbf{Metrics}} \\ \cline{3-11}
     &  & {NDCG@5}  & {NDCG@10}  & {NDCG@20} & {HR@5}  & 
    {HR@10}  & {HR@20}  & {Precision@5} & {Precision@10} & {Precision@20}  \\
    \midrule
    \midrule
    \multirow{2}{*}{\texttt{Beauty}} & {20} 
    & \textbf{0.0097} & 0.0124 & 0.0143 & \textbf{0.0152} & 0.0238 & 0.0312 & \textbf{0.0030} & 0.0024 & 0.0016 \\
    & {50} 
    & 0.0094 & \textbf{0.0126} & \textbf{0.0154} & 0.0150 & \textbf{0.0248} & \textbf{0.0361} & \textbf{0.0030} & \textbf{0.0025} & \textbf{0.0018} \\
    \hline
    \multirow{2}{*}{\texttt{Toys}} & {20} 
    &  \textbf{0.0067} & 0.0080 & 0.0090 & \textbf{0.0106} & 0.0148 & 0.0186 & \textbf{0.0021} & 0.0015 & 0.0009 \\   
    & {50} 
    & 0.0066 & \textbf{0.0084} & \textbf{0.0096} & 0.0105 & \textbf{0.0158} & \textbf{0.0209} & \textbf{0.0021} & \textbf{0.0016} & \textbf{0.0010} \\
    \hline
    \multirow{2}{*}{\texttt{Yelp}} & {20} 
    & \textbf{0.0099} & \textbf{0.0121} & 0.0136 & \textbf{0.0148} & \textbf{0.0214} & 0.0276 & \textbf{0.0030} & \textbf{0.0021} & 0.0014 \\
    & {50} 
    & 0.0095 & 0.0116 & \textbf{0.0144} & 0.0145 & 0.0213 & \textbf{0.0323} & 0.0029 & \textbf{0.0021} & \textbf{0.0016} \\
    \bottomrule[2pt]
    \end{tabular}
    }
    \vspace{-0.2cm}
\end{table*}

\subsubsection{The impact of collaborative Inference.}
    To analyze the effectiveness of each component in our collaborative inference strategy, we conducted ablation experiments on the inference strategy. As shown in Table~\ref{tab:ablation_inference}, each dataset has four rows of experimental results, corresponding to four types of ablations. For the four rows of data for each dataset, we provide the following explanations:
    \begin{itemize}[leftmargin=*]
        \item \textbf{w.} LLM, \textbf{w/o.} SRM, \textbf{w/o.} Result Fusion (Row.1) indicates that only use on-cloud LLM to recall the candidate list, the initial ranking results can be regarded as the full-ranking results based on the less real-time data.
        \item \textbf{w/o.} LLM, \textbf{w.} SRM, \textbf{w/o.} Result Fusion (Row.2) indicates that only use on-device SRM to do ranking.
        \item \textbf{w.} LLM, \textbf{w.} SRM, \textbf{w/o.} Result Fusion (Row.3) indicates first to let on-cloud LLM recall a candidate list, then on-device SRM do rerank the candidate list based on the real-time
        \item \textbf{w.} LLM, \textbf{w.} SRM, \textbf{w.} Result Fusion (Row.4) indicates first to let on-cloud LLM recall a candidate list and initial ranking resuls, then on-device SRM do rerank the candidate list based on the real-time, further, reranking and initial ranking ard fused as the final ranking results.
    \end{itemize}
The experimental results show that the collaborative inference and result fusion of LLMs and SRMs perform the best, followed by collaborative inference of LLMs and SRMs without result fusion. Using either the LLM or the SRM alone results in poorer performance. The above results demonstrate the effectiveness of our designed collaborative inference and result fusion.

\subsubsection{The impact of collaborative-decision request.}
As shown in Table~\ref{tab:ablation_request}, we analyze the collaborative-decision request.

\begin{itemize}[leftmargin=*]
    \item \textbf{Random} (Row.1) indicates that choosing some users to update the candidate list and corresponded initial ranking results randomly.
    \item \textbf{Inconsistency} (Row.2) indicates that choosing some users to update the candidate list and corresponded initial ranking results based on the ranking results of the LLM and SRM. High inconsistency can get more priority to update.
\end{itemize}

The experimental results indicate that our strategy of selecting user data for generating new candidate lists and initial ranking results with LLM, based on the inconsistency of ranking between LLM and SRM, achieves an average improvement of about 25\% compared to random selection. This is because our strategy tends to prioritize updates of candidate lists and initial rankings with the highest benefit, while random updating lacks the capability to calculate the benefits of updates. At the same time, we can observe that the peak of improvement occurs at a request frequency of 10\%, and then as the frequency increases, the magnitude of improvement diminishes. This is because when the request frequency increases (e.g., to 20\%, 40\%), our strategy has already updated the parts with the highest benefits, while random updates mix high and low benefit updates, leading to a decrease in improvement. In summary, the above results all demonstrate that our strategy effectively selects the users with the highest benefits for updating candidate lists and initial rankings.

\subsection{In-Depth Analysis}
\subsubsection{The impact of length of candidate list.}
In Table~\ref{tab:analysis_candidate_length}, we analyze the impact of the length of the LLM's recall list on performance. When the length of the candidate list is 20 and 50, the range of performance variation is not significant, which indicates that LSC4Rec is not very sensitive to this hyperparameter of the recall list's length. 
Additionally, shorter candidate lists tend to perform better on stricter metrics (e.g., @5), while longer candidate lists show advantages on more lenient metrics (e.g., @20). The advantage of a longer candidate list lies in its higher likelihood of including the ground-truth item, but the downside is the greater challenge for the SRM to re-rank longer candidate lists effectively. On a relatively balanced metric (e.g., @10), the advantages and disadvantages of longer candidate lists are approximately equal.

\begin{table}[!ht]
\caption{Resource consumption comparison.}
\label{tab:consumption}
\vspace{-0.3cm}
\resizebox{0.4\textwidth}{!}{
\begin{tabular}{c|c|c|c}
\toprule[2pt]
\textbf{Model} & \textbf{\#Parameter} & \begin{tabular}[c]{@{}c@{}}\textbf{Training Time}\\      \textbf{(s/epoch)}\end{tabular} & \begin{tabular}[c]{@{}c@{}}\textbf{Inference Time}\\      \textbf{(s/sample)}\end{tabular} \\
\midrule \midrule
\rowcolor{blue!3}\multicolumn{4}{c}{\textit{On-device}} \\
\midrule
\rowcolor{blue!3}DIN & 0.82M & 38.50 & 0.00074 \\
\hline
\rowcolor{blue!3}GRU4Rec & 0.81M & 38.30 & 0.00085 \\
\hline
\rowcolor{blue!3}SASRec & 1.62M & 39.40 & 0.00143 \\
\midrule
\rowcolor{green!3}\multicolumn{4}{c}{\textit{On-cloud}} \\
\midrule
\rowcolor{green!3}P5 & 60.75M & 897.83 & 0.10082 \\
\hline
\rowcolor{green!3}POD & 60.77M & 2277.16 & 0.32563 \\
\bottomrule[2pt]
\end{tabular}
}
\vspace{-0.35cm}
\end{table}
\subsubsection{Analysis of resource consumption}
To validate the resource consumption of our method, we analyzed the resource usage of our approach in Table~\ref{tab:consumption}. Since the choice of either P5 or POD as the LLM does not affect the conclusion, we will use one of the models as an example. Here, we choose P5 to describe the conclusion.
In terms of the number of parameters, if we uses P5 as LLM and SASRec as SRM, the total number of parameters is the sum of P5 and SASRec (on-device consumption is same as SASRec). In terms of training duration, since our method does not require real-time training of LLMs, the time consumption is reduced compared to training LLMs. In terms of inference duration, because our method does not need to call the LLM at every moment, and often does not need to call the LLM at all, the time consumption is significantly reduced compared to the LLM, and has not increased much compared to the SRM, which does not affect the real-time nature of the recommendations. Overall, our method leverages the advantages of LLMs with low resource consumption to achieve better performance.
\section{Conclusion}
\label{sec:conclusion}

The proposed Device-Cloud Collaborative Framework for LLM and SRM (LSC4Rec) effectively combines the strengths of large language models and small recommendation models. By leveraging collaborative training, collaborative inference, and intelligent request strategies, LSC4Rec addresses the limitations of LLMs in capturing real-time user preferences while optimizing the utilization of cloud and edge computing resources. Extensive experimental results validate the effectiveness of each strategy, highlighting the framework's potential for practical applications. Future work will focus on further refining the collaboration mechanisms to enhance recommendation performance and user satisfaction.
\section*{ACKNOWLEDGEMENTS}
\begin{sloppypar}
This work was supported by 2030 National Science and Technology Major Project (2022ZD0119100), Scientific Research Fund of Zhejiang Provincial Education Department (Y202353679), National Natural Science Foundation of China (No. 62402429, 62376243, 62441605, 62037001), the Key Research and Development Program of Zhejiang Province (No. 2024C03270), ZJU Kunpeng$\&$Ascend Center of Excellence, the Key Research and Development Projects in Zhejiang Province (No.2024C01106), Zhejiang University Education Foundation Qizhen Scholar Foundation, the Starry Night Science Fund at Shanghai Institute for Advanced Study (Zhejiang University). This work was also supported by Ant group.
\end{sloppypar}

\bibliographystyle{ACM-Reference-Format}
\balance
\bibliography{reference}


\begin{thebibliography}{66}


\ifx \showCODEN    \undefined \def \showCODEN     #1{\unskip}     \fi
\ifx \showISBNx    \undefined \def \showISBNx     #1{\unskip}     \fi
\ifx \showISBNxiii \undefined \def \showISBNxiii  #1{\unskip}     \fi
\ifx \showISSN     \undefined \def \showISSN      #1{\unskip}     \fi
\ifx \showLCCN     \undefined \def \showLCCN      #1{\unskip}     \fi
\ifx \shownote     \undefined \def \shownote      #1{#1}          \fi
\ifx \showarticletitle \undefined \def \showarticletitle #1{#1}   \fi
\ifx \showURL      \undefined \def \showURL       {\relax}        \fi
\providecommand\bibfield[2]{#2}
\providecommand\bibinfo[2]{#2}
\providecommand\natexlab[1]{#1}
\providecommand\showeprint[2][]{arXiv:#2}

\bibitem[Bao et~al\mbox{.}(2023a)]%
        {ref:llm4rec_Tallrec}
\bibfield{author}{\bibinfo{person}{Keqin Bao}, \bibinfo{person}{Jizhi Zhang}, \bibinfo{person}{Yang Zhang}, \bibinfo{person}{Wenjie Wang}, \bibinfo{person}{Fuli Feng}, {and} \bibinfo{person}{Xiangnan He}.} \bibinfo{year}{2023}\natexlab{a}.
\newblock \showarticletitle{TALLRec: An Effective and Efficient Tuning Framework to Align Large Language Model with Recommendation}.
\newblock  (\bibinfo{year}{2023}), \bibinfo{pages}{1007--1014}.
\newblock


\bibitem[Bao et~al\mbox{.}(2023b)]%
        {bao2023tallrec}
\bibfield{author}{\bibinfo{person}{Keqin Bao}, \bibinfo{person}{Jizhi Zhang}, \bibinfo{person}{Yang Zhang}, \bibinfo{person}{Wenjie Wang}, \bibinfo{person}{Fuli Feng}, {and} \bibinfo{person}{Xiangnan He}.} \bibinfo{year}{2023}\natexlab{b}.
\newblock \showarticletitle{Tallrec: An effective and efficient tuning framework to align large language model with recommendation}. In \bibinfo{booktitle}{\emph{Proceedings of the 17th ACM Conference on Recommender Systems}}. \bibinfo{publisher}{{ACM}}, \bibinfo{pages}{1007--1014}.
\newblock


\bibitem[Chen et~al\mbox{.}(2022)]%
        {chen2022intent}
\bibfield{author}{\bibinfo{person}{Yongjun Chen}, \bibinfo{person}{Zhiwei Liu}, \bibinfo{person}{Jia Li}, \bibinfo{person}{Julian McAuley}, {and} \bibinfo{person}{Caiming Xiong}.} \bibinfo{year}{2022}\natexlab{}.
\newblock \showarticletitle{Intent contrastive learning for sequential recommendation}. In \bibinfo{booktitle}{\emph{Proceedings of the ACM Web Conference 2022}}. \bibinfo{pages}{2172--2182}.
\newblock


\bibitem[Chen et~al\mbox{.}(2021a)]%
        {chen2021improving}
\bibfield{author}{\bibinfo{person}{Zhengyu Chen}, \bibinfo{person}{Donglin Wang}, {and} \bibinfo{person}{Shiqian Yin}.} \bibinfo{year}{2021}\natexlab{a}.
\newblock \showarticletitle{Improving cold-start recommendation via multi-prior meta-learning}. In \bibinfo{booktitle}{\emph{Advances in Information Retrieval: 43rd European Conference on IR Research, ECIR 2021, Virtual Event, March 28--April 1, 2021, Proceedings, Part II 43}}. Springer, \bibinfo{pages}{249--256}.
\newblock


\bibitem[Chen et~al\mbox{.}(2021b)]%
        {chen2021deep}
\bibfield{author}{\bibinfo{person}{Zhengyu Chen}, \bibinfo{person}{Ziqing Xu}, {and} \bibinfo{person}{Donglin Wang}.} \bibinfo{year}{2021}\natexlab{b}.
\newblock \showarticletitle{Deep transfer tensor decomposition with orthogonal constraint for recommender systems}. In \bibinfo{booktitle}{\emph{Proceedings of the AAAI Conference on Artificial Intelligence}}, Vol.~\bibinfo{volume}{35}. \bibinfo{pages}{4010--4018}.
\newblock


\bibitem[Dai et~al\mbox{.}(2023)]%
        {dai2023uncovering}
\bibfield{author}{\bibinfo{person}{Sunhao Dai}, \bibinfo{person}{Ninglu Shao}, \bibinfo{person}{Haiyuan Zhao}, \bibinfo{person}{Weijie Yu}, \bibinfo{person}{Zihua Si}, \bibinfo{person}{Chen Xu}, \bibinfo{person}{Zhongxiang Sun}, \bibinfo{person}{Xiao Zhang}, {and} \bibinfo{person}{Jun Xu}.} \bibinfo{year}{2023}\natexlab{}.
\newblock \showarticletitle{Uncovering ChatGPT's Capabilities in Recommender Systems}. In \bibinfo{booktitle}{\emph{RecSys}}. \bibinfo{publisher}{{ACM}}, \bibinfo{pages}{1126--1132}.
\newblock


\bibitem[Fu et~al\mbox{.}(2023)]%
        {fu2023end}
\bibfield{author}{\bibinfo{person}{Kairui Fu}, \bibinfo{person}{Qiaowei Miao}, \bibinfo{person}{Shengyu Zhang}, \bibinfo{person}{Kun Kuang}, {and} \bibinfo{person}{Fei Wu}.} \bibinfo{year}{2023}\natexlab{}.
\newblock \showarticletitle{End-to-End Optimization of Quantization-Based Structure Learning and Interventional Next-Item Recommendation}. In \bibinfo{booktitle}{\emph{CAAI International Conference on Artificial Intelligence}}. Springer, \bibinfo{pages}{415--429}.
\newblock


\bibitem[Fu et~al\mbox{.}(2024)]%
        {ref:diet}
\bibfield{author}{\bibinfo{person}{Kairui Fu}, \bibinfo{person}{Shengyu Zhang}, \bibinfo{person}{Zheqi Lv}, \bibinfo{person}{Jingyuan Chen}, {and} \bibinfo{person}{Jiwei Li}.} \bibinfo{year}{2024}\natexlab{}.
\newblock \showarticletitle{{DIET:} Customized Slimming for Incompatible Networks in Sequential Recommendation}. In \bibinfo{booktitle}{\emph{{KDD}}}. \bibinfo{publisher}{{ACM}}, \bibinfo{pages}{816--826}.
\newblock


\bibitem[Gao et~al\mbox{.}(2023b)]%
        {gao2023cirs}
\bibfield{author}{\bibinfo{person}{Chongming Gao}, \bibinfo{person}{Shiqi Wang}, \bibinfo{person}{Shijun Li}, \bibinfo{person}{Jiawei Chen}, \bibinfo{person}{Xiangnan He}, \bibinfo{person}{Wenqiang Lei}, \bibinfo{person}{Biao Li}, \bibinfo{person}{Yuan Zhang}, {and} \bibinfo{person}{Peng Jiang}.} \bibinfo{year}{2023}\natexlab{b}.
\newblock \showarticletitle{CIRS: Bursting Filter Bubbles by Counterfactual Interactive Recommender System}.
\newblock \bibinfo{journal}{\emph{ACM Transactions on Information Systems (TOIS)}} \bibinfo{volume}{42}, \bibinfo{number}{1}, Article \bibinfo{articleno}{14} (\bibinfo{date}{aug} \bibinfo{year}{2023}), \bibinfo{numpages}{27}~pages.
\newblock
\showISSN{1046-8188}
\href{https://doi.org/10.1145/3594871}{doi:\nolinkurl{10.1145/3594871}}


\bibitem[Gao et~al\mbox{.}(2023a)]%
        {ref:llm_chatrec}
\bibfield{author}{\bibinfo{person}{Yunfan Gao}, \bibinfo{person}{Tao Sheng}, \bibinfo{person}{Youlin Xiang}, \bibinfo{person}{Yun Xiong}, \bibinfo{person}{Haofen Wang}, {and} \bibinfo{person}{Jiawei Zhang}.} \bibinfo{year}{2023}\natexlab{a}.
\newblock \showarticletitle{Chat-rec: Towards interactive and explainable llms-augmented recommender system}.
\newblock \bibinfo{journal}{\emph{arXiv preprint arXiv:2303.14524}} (\bibinfo{year}{2023}).
\newblock


\bibitem[Geng et~al\mbox{.}(2022)]%
        {ref:llm4rec_P5}
\bibfield{author}{\bibinfo{person}{Shijie Geng}, \bibinfo{person}{Shuchang Liu}, \bibinfo{person}{Zuohui Fu}, \bibinfo{person}{Yingqiang Ge}, {and} \bibinfo{person}{Yongfeng Zhang}.} \bibinfo{year}{2022}\natexlab{}.
\newblock \showarticletitle{Recommendation as Language Processing {(RLP):} {A} Unified Pretrain, Personalized Prompt {\&} Predict Paradigm {(P5)}}. In \bibinfo{booktitle}{\emph{RecSys}}. \bibinfo{publisher}{{ACM}}, \bibinfo{pages}{299--315}.
\newblock


\bibitem[Hidasi et~al\mbox{.}(2016)]%
        {ref:gru4rec}
\bibfield{author}{\bibinfo{person}{Bal{\'a}zs Hidasi}, \bibinfo{person}{Alexandros Karatzoglou}, \bibinfo{person}{Linas Baltrunas}, {and} \bibinfo{person}{Domonkos Tikk}.} \bibinfo{year}{2016}\natexlab{}.
\newblock \showarticletitle{Session-based recommendations with recurrent neural networks}.
\newblock \bibinfo{journal}{\emph{International Conference on Learning Representations 2016}} (\bibinfo{year}{2016}).
\newblock


\bibitem[Ji et~al\mbox{.}(2023a)]%
        {ref:genrec}
\bibfield{author}{\bibinfo{person}{Jianchao Ji}, \bibinfo{person}{Zelong Li}, \bibinfo{person}{Shuyuan Xu}, \bibinfo{person}{Wenyue Hua}, \bibinfo{person}{Yingqiang Ge}, \bibinfo{person}{Juntao Tan}, {and} \bibinfo{person}{Yongfeng Zhang}.} \bibinfo{year}{2023}\natexlab{a}.
\newblock \showarticletitle{Genrec: Large language model for generative recommendation}.
\newblock \bibinfo{journal}{\emph{arXiv e-prints}} (\bibinfo{year}{2023}), \bibinfo{pages}{arXiv--2307}.
\newblock


\bibitem[Ji et~al\mbox{.}(2023b)]%
        {ji2023online}
\bibfield{author}{\bibinfo{person}{Wei Ji}, \bibinfo{person}{Xiangyan Liu}, \bibinfo{person}{An Zhang}, \bibinfo{person}{Yinwei Wei}, \bibinfo{person}{Yongxin Ni}, {and} \bibinfo{person}{Xiang Wang}.} \bibinfo{year}{2023}\natexlab{b}.
\newblock \showarticletitle{Online distillation-enhanced multi-modal transformer for sequential recommendation}. In \bibinfo{booktitle}{\emph{Proceedings of the 31st ACM International Conference on Multimedia}}. \bibinfo{pages}{955--965}.
\newblock


\bibitem[Kang and McAuley(2018)]%
        {ref:sasrec}
\bibfield{author}{\bibinfo{person}{Wang{-}Cheng Kang} {and} \bibinfo{person}{Julian~J. McAuley}.} \bibinfo{year}{2018}\natexlab{}.
\newblock \showarticletitle{Self-Attentive Sequential Recommendation}. In \bibinfo{booktitle}{\emph{{IEEE} International Conference on Data Mining, {ICDM} 2018, Singapore, November 17-20, 2018}}. \bibinfo{publisher}{{IEEE} Computer Society}, \bibinfo{pages}{197--206}.
\newblock
\href{https://doi.org/10.1109/ICDM.2018.00035}{doi:\nolinkurl{10.1109/ICDM.2018.00035}}


\bibitem[Li et~al\mbox{.}(2025)]%
        {li_llm4code}
\bibfield{author}{\bibinfo{person}{Heng Li}, \bibinfo{person}{Zhiyuan Yao}, \bibinfo{person}{Bang Wu}, \bibinfo{person}{Cuiying Gao}, \bibinfo{person}{Teng Xu}, \bibinfo{person}{Wei Yuan}, {and} \bibinfo{person}{Xiapu Luo}.} \bibinfo{year}{2025}\natexlab{}.
\newblock \showarticletitle{Automated Malware Assembly Line: Uniting Piggybacking and Adversarial Example in Android Malware Generation}. In \bibinfo{booktitle}{\emph{32nd Annual Network and Distributed System Security Symposium, {NDSS} 2025, San Diego, California, USA, February 24 - February 28, 2025}}. \bibinfo{publisher}{The Internet Society}.
\newblock


\bibitem[Li et~al\mbox{.}(2023)]%
        {ref:llm4rec_Prompt_Distillation}
\bibfield{author}{\bibinfo{person}{Lei Li}, \bibinfo{person}{Yongfeng Zhang}, {and} \bibinfo{person}{Li Chen}.} \bibinfo{year}{2023}\natexlab{}.
\newblock \showarticletitle{Prompt Distillation for Efficient LLM-based Recommendation}. In \bibinfo{booktitle}{\emph{{CIKM}}}. \bibinfo{publisher}{{ACM}}, \bibinfo{pages}{1348--1357}.
\newblock


\bibitem[Li et~al\mbox{.}(2024)]%
        {li2024tokenpacker}
\bibfield{author}{\bibinfo{person}{Wentong Li}, \bibinfo{person}{Yuqian Yuan}, \bibinfo{person}{Jian Liu}, \bibinfo{person}{Dongqi Tang}, \bibinfo{person}{Song Wang}, \bibinfo{person}{Jie Qin}, \bibinfo{person}{Jianke Zhu}, {and} \bibinfo{person}{Lei Zhang}.} \bibinfo{year}{2024}\natexlab{}.
\newblock \showarticletitle{Tokenpacker: Efficient visual projector for multimodal llm}.
\newblock \bibinfo{journal}{\emph{arXiv preprint arXiv:2407.02392}} (\bibinfo{year}{2024}).
\newblock


\bibitem[Li et~al\mbox{.}(2022)]%
        {li2022corec}
\bibfield{author}{\bibinfo{person}{Yangfan Li}, \bibinfo{person}{Kenli Li}, \bibinfo{person}{Wei Wei}, \bibinfo{person}{Tianyi Zhou}, {and} \bibinfo{person}{Cen Chen}.} \bibinfo{year}{2022}\natexlab{}.
\newblock \showarticletitle{CoRec: an efficient internet behavior-based recommendation framework with edge-cloud collaboration on deep convolution neural networks}.
\newblock \bibinfo{journal}{\emph{ACM Transactions on Sensor Networks}} \bibinfo{volume}{19}, \bibinfo{number}{2} (\bibinfo{year}{2022}), \bibinfo{pages}{1--28}.
\newblock


\bibitem[Liao et~al\mbox{.}(2023)]%
        {liao2023ppgencdr}
\bibfield{author}{\bibinfo{person}{Xinting Liao}, \bibinfo{person}{Weiming Liu}, \bibinfo{person}{Xiaolin Zheng}, \bibinfo{person}{Binhui Yao}, {and} \bibinfo{person}{Chaochao Chen}.} \bibinfo{year}{2023}\natexlab{}.
\newblock \showarticletitle{Ppgencdr: A stable and robust framework for privacy-preserving cross-domain recommendation}. In \bibinfo{booktitle}{\emph{Proceedings of the AAAI Conference on Artificial Intelligence}}, Vol.~\bibinfo{volume}{37}. \bibinfo{pages}{4453--4461}.
\newblock


\bibitem[Lightman et~al\mbox{.}(2023)]%
        {lightman2023let}
\bibfield{author}{\bibinfo{person}{Hunter Lightman}, \bibinfo{person}{Vineet Kosaraju}, \bibinfo{person}{Yura Burda}, \bibinfo{person}{Harri Edwards}, \bibinfo{person}{Bowen Baker}, \bibinfo{person}{Teddy Lee}, \bibinfo{person}{Jan Leike}, \bibinfo{person}{John Schulman}, \bibinfo{person}{Ilya Sutskever}, {and} \bibinfo{person}{Karl Cobbe}.} \bibinfo{year}{2023}\natexlab{}.
\newblock \showarticletitle{Let's verify step by step}.
\newblock \bibinfo{journal}{\emph{arXiv preprint arXiv:2305.20050}} (\bibinfo{year}{2023}).
\newblock


\bibitem[Lin et~al\mbox{.}(2024a)]%
        {lin2024bridging}
\bibfield{author}{\bibinfo{person}{Xinyu Lin}, \bibinfo{person}{Wenjie Wang}, \bibinfo{person}{Yongqi Li}, \bibinfo{person}{Fuli Feng}, \bibinfo{person}{See-Kiong Ng}, {and} \bibinfo{person}{Tat-Seng Chua}.} \bibinfo{year}{2024}\natexlab{a}.
\newblock \showarticletitle{Bridging items and language: A transition paradigm for large language model-based recommendation}. In \bibinfo{booktitle}{\emph{Proceedings of the 30th ACM SIGKDD Conference on Knowledge Discovery and Data Mining}}. \bibinfo{pages}{1816--1826}.
\newblock


\bibitem[Lin et~al\mbox{.}(2024b)]%
        {lin2024efficient}
\bibfield{author}{\bibinfo{person}{Xinyu Lin}, \bibinfo{person}{Chaoqun Yang}, \bibinfo{person}{Wenjie Wang}, \bibinfo{person}{Yongqi Li}, \bibinfo{person}{Cunxiao Du}, \bibinfo{person}{Fuli Feng}, \bibinfo{person}{See{-}Kiong Ng}, {and} \bibinfo{person}{Tat{-}Seng Chua}.} \bibinfo{year}{2024}\natexlab{b}.
\newblock \showarticletitle{Efficient Inference for Large Language Model-based Generative Recommendation}.
\newblock \bibinfo{journal}{\emph{CoRR}}  \bibinfo{volume}{abs/2410.05165} (\bibinfo{year}{2024}).
\newblock


\bibitem[Liu et~al\mbox{.}(2023a)]%
        {ref:llm_rec_14}
\bibfield{author}{\bibinfo{person}{Junling Liu}, \bibinfo{person}{Chao Liu}, \bibinfo{person}{Renjie Lv}, \bibinfo{person}{Kang Zhou}, {and} \bibinfo{person}{Yan Zhang}.} \bibinfo{year}{2023}\natexlab{a}.
\newblock \showarticletitle{Is ChatGPT a Good Recommender? {A} Preliminary Study}.
\newblock \bibinfo{journal}{\emph{CoRR}}  \bibinfo{volume}{abs/2304.10149} (\bibinfo{year}{2023}).
\newblock


\bibitem[Liu et~al\mbox{.}(2024a)]%
        {liu2024educating}
\bibfield{author}{\bibinfo{person}{Kai Liu}, \bibinfo{person}{Ze Chen}, \bibinfo{person}{Zhihang Fu}, \bibinfo{person}{Rongxin Jiang}, \bibinfo{person}{Fan Zhou}, \bibinfo{person}{Yaowu Chen}, \bibinfo{person}{Yue Wu}, {and} \bibinfo{person}{Jieping Ye}.} \bibinfo{year}{2024}\natexlab{a}.
\newblock \showarticletitle{Structure-aware Domain Knowledge Injection for Large Language Models}.
\newblock \bibinfo{journal}{\emph{arXiv preprint arXiv:2407.16724}} (\bibinfo{year}{2024}).
\newblock


\bibitem[Liu et~al\mbox{.}(2024c)]%
        {liu2024enhancing}
\bibfield{author}{\bibinfo{person}{Kai Liu}, \bibinfo{person}{Zhihang Fu}, \bibinfo{person}{Chao Chen}, \bibinfo{person}{Wei Zhang}, \bibinfo{person}{Rongxin Jiang}, \bibinfo{person}{Fan Zhou}, \bibinfo{person}{Yaowu Chen}, \bibinfo{person}{Yue Wu}, {and} \bibinfo{person}{Jieping Ye}.} \bibinfo{year}{2024}\natexlab{c}.
\newblock \showarticletitle{Enhancing LLM's Cognition via Structurization}.
\newblock \bibinfo{journal}{\emph{Advances in Neural Information Processing Systems}}  \bibinfo{volume}{38} (\bibinfo{year}{2024}).
\newblock


\bibitem[Liu et~al\mbox{.}(2024b)]%
        {liu2024learning}
\bibfield{author}{\bibinfo{person}{Weiming Liu}, \bibinfo{person}{Chaochao Chen}, \bibinfo{person}{Xinting Liao}, \bibinfo{person}{Mengling Hu}, \bibinfo{person}{Yanchao Tan}, \bibinfo{person}{Fan Wang}, \bibinfo{person}{Xiaolin Zheng}, {and} \bibinfo{person}{Yew~Soon Ong}.} \bibinfo{year}{2024}\natexlab{b}.
\newblock \showarticletitle{Learning Accurate and Bidirectional Transformation via Dynamic Embedding Transportation for Cross-Domain Recommendation}. In \bibinfo{booktitle}{\emph{Proceedings of the AAAI Conference on Artificial Intelligence}}, Vol.~\bibinfo{volume}{38}. \bibinfo{pages}{8815--8823}.
\newblock


\bibitem[Liu et~al\mbox{.}(2023b)]%
        {liu2023joint}
\bibfield{author}{\bibinfo{person}{Weiming Liu}, \bibinfo{person}{Xiaolin Zheng}, \bibinfo{person}{Chaochao Chen}, \bibinfo{person}{Jiajie Su}, \bibinfo{person}{Xinting Liao}, \bibinfo{person}{Mengling Hu}, {and} \bibinfo{person}{Yanchao Tan}.} \bibinfo{year}{2023}\natexlab{b}.
\newblock \showarticletitle{Joint internal multi-interest exploration and external domain alignment for cross domain sequential recommendation}. In \bibinfo{booktitle}{\emph{Proceedings of the ACM Web Conference 2023}}. \bibinfo{pages}{383--394}.
\newblock


\bibitem[Long et~al\mbox{.}(2024a)]%
        {ref:device_cloud_rec}
\bibfield{author}{\bibinfo{person}{Jing Long}, \bibinfo{person}{Guanhua Ye}, \bibinfo{person}{Tong Chen}, \bibinfo{person}{Yang Wang}, \bibinfo{person}{Meng Wang}, {and} \bibinfo{person}{Hongzhi Yin}.} \bibinfo{year}{2024}\natexlab{a}.
\newblock \showarticletitle{Diffusion-Based Cloud-Edge-Device Collaborative Learning for Next {POI} Recommendations}. In \bibinfo{booktitle}{\emph{{KDD}}}. \bibinfo{publisher}{{ACM}}, \bibinfo{pages}{2026--2036}.
\newblock


\bibitem[Long et~al\mbox{.}(2024b)]%
        {long2024diffusion}
\bibfield{author}{\bibinfo{person}{Jing Long}, \bibinfo{person}{Guanhua Ye}, \bibinfo{person}{Tong Chen}, \bibinfo{person}{Yang Wang}, \bibinfo{person}{Meng Wang}, {and} \bibinfo{person}{Hongzhi Yin}.} \bibinfo{year}{2024}\natexlab{b}.
\newblock \showarticletitle{Diffusion-based cloud-edge-device collaborative learning for next POI recommendations}. In \bibinfo{booktitle}{\emph{Proceedings of the 30th ACM SIGKDD Conference on Knowledge Discovery and Data Mining}}. \bibinfo{pages}{2026--2036}.
\newblock


\bibitem[Luo et~al\mbox{.}(2023)]%
        {rec:recranker}
\bibfield{author}{\bibinfo{person}{Sichun Luo}, \bibinfo{person}{Bowei He}, \bibinfo{person}{Haohan Zhao}, \bibinfo{person}{Yinya Huang}, \bibinfo{person}{Aojun Zhou}, \bibinfo{person}{Zongpeng Li}, \bibinfo{person}{Yuanzhang Xiao}, \bibinfo{person}{Mingjie Zhan}, {and} \bibinfo{person}{Linqi Song}.} \bibinfo{year}{2023}\natexlab{}.
\newblock \showarticletitle{RecRanker: Instruction Tuning Large Language Model as Ranker for Top-k Recommendation}.
\newblock \bibinfo{journal}{\emph{CoRR}}  \bibinfo{volume}{abs/2312.16018} (\bibinfo{year}{2023}).
\newblock


\bibitem[Lv et~al\mbox{.}(2024a)]%
        {lv2024semantic}
\bibfield{author}{\bibinfo{person}{Zheqi Lv}, \bibinfo{person}{Shaoxuan He}, \bibinfo{person}{Tianyu Zhan}, \bibinfo{person}{Shengyu Zhang}, \bibinfo{person}{Wenqiao Zhang}, \bibinfo{person}{Jingyuan Chen}, \bibinfo{person}{Zhou Zhao}, {and} \bibinfo{person}{Fei Wu}.} \bibinfo{year}{2024}\natexlab{a}.
\newblock \showarticletitle{Semantic Codebook Learning for Dynamic Recommendation Models}. In \bibinfo{booktitle}{\emph{{ACM} Multimedia}}. \bibinfo{publisher}{{ACM}}, \bibinfo{pages}{9611--9620}.
\newblock


\bibitem[Lv et~al\mbox{.}(2024b)]%
        {lv2024intelligent}
\bibfield{author}{\bibinfo{person}{Zheqi Lv}, \bibinfo{person}{Wenqiao Zhang}, \bibinfo{person}{Zhengyu Chen}, \bibinfo{person}{Shengyu Zhang}, {and} \bibinfo{person}{Kun Kuang}.} \bibinfo{year}{2024}\natexlab{b}.
\newblock \showarticletitle{Intelligent Model Update Strategy for Sequential Recommendation}. In \bibinfo{booktitle}{\emph{{WWW}}}. \bibinfo{publisher}{{ACM}}, \bibinfo{pages}{3117--3128}.
\newblock


\bibitem[Lv et~al\mbox{.}(2023)]%
        {lv2023duet}
\bibfield{author}{\bibinfo{person}{Zheqi Lv}, \bibinfo{person}{Wenqiao Zhang}, \bibinfo{person}{Shengyu Zhang}, \bibinfo{person}{Kun Kuang}, \bibinfo{person}{Feng Wang}, \bibinfo{person}{Yongwei Wang}, \bibinfo{person}{Zhengyu Chen}, \bibinfo{person}{Tao Shen}, \bibinfo{person}{Hongxia Yang}, \bibinfo{person}{Beng~Chin Ooi}, {and} \bibinfo{person}{Fei Wu}.} \bibinfo{year}{2023}\natexlab{}.
\newblock \showarticletitle{{DUET:} {A} Tuning-Free Device-Cloud Collaborative Parameters Generation Framework for Efficient Device Model Generalization}. In \bibinfo{booktitle}{\emph{{WWW}}}. \bibinfo{publisher}{{ACM}}, \bibinfo{pages}{3077--3085}.
\newblock


\bibitem[Qian et~al\mbox{.}(2022)]%
        {ref:device_cloud_adarequest}
\bibfield{author}{\bibinfo{person}{Xufeng Qian}, \bibinfo{person}{Yue Xu}, \bibinfo{person}{Fuyu Lv}, \bibinfo{person}{Shengyu Zhang}, \bibinfo{person}{Ziwen Jiang}, \bibinfo{person}{Qingwen Liu}, \bibinfo{person}{Xiaoyi Zeng}, \bibinfo{person}{Tat-Seng Chua}, {and} \bibinfo{person}{Fei Wu}.} \bibinfo{year}{2022}\natexlab{}.
\newblock \showarticletitle{Intelligent request strategy design in recommender system}. In \bibinfo{booktitle}{\emph{Proceedings of the 28th ACM SIGKDD Conference on Knowledge Discovery and Data Mining}}. \bibinfo{pages}{3772--3782}.
\newblock


\bibitem[Raffel et~al\mbox{.}(2020)]%
        {ref:unified_trans}
\bibfield{author}{\bibinfo{person}{Colin Raffel}, \bibinfo{person}{Noam Shazeer}, \bibinfo{person}{Adam Roberts}, \bibinfo{person}{Katherine Lee}, \bibinfo{person}{Sharan Narang}, \bibinfo{person}{Michael Matena}, \bibinfo{person}{Yanqi Zhou}, \bibinfo{person}{Wei Li}, {and} \bibinfo{person}{Peter~J Liu}.} \bibinfo{year}{2020}\natexlab{}.
\newblock \showarticletitle{Exploring the limits of transfer learning with a unified text-to-text transformer}.
\newblock \bibinfo{journal}{\emph{The Journal of Machine Learning Research}} \bibinfo{volume}{21}, \bibinfo{number}{1} (\bibinfo{year}{2020}), \bibinfo{pages}{5485--5551}.
\newblock


\bibitem[Su et~al\mbox{.}(2023)]%
        {su2023enhancing}
\bibfield{author}{\bibinfo{person}{Jiajie Su}, \bibinfo{person}{Chaochao Chen}, \bibinfo{person}{Weiming Liu}, \bibinfo{person}{Fei Wu}, \bibinfo{person}{Xiaolin Zheng}, {and} \bibinfo{person}{Haoming Lyu}.} \bibinfo{year}{2023}\natexlab{}.
\newblock \showarticletitle{Enhancing hierarchy-aware graph networks with deep dual clustering for session-based recommendation}. In \bibinfo{booktitle}{\emph{Proceedings of the ACM Web Conference 2023}}. \bibinfo{pages}{165--176}.
\newblock


\bibitem[Sun et~al\mbox{.}(2022)]%
        {sun2022response}
\bibfield{author}{\bibinfo{person}{Teng Sun}, \bibinfo{person}{Chun Wang}, \bibinfo{person}{Xuemeng Song}, \bibinfo{person}{Fuli Feng}, {and} \bibinfo{person}{Liqiang Nie}.} \bibinfo{year}{2022}\natexlab{}.
\newblock \showarticletitle{Response generation by jointly modeling personalized linguistic styles and emotions}.
\newblock \bibinfo{journal}{\emph{ACM Transactions on Multimedia Computing, Communications, and Applications (TOMM)}} \bibinfo{volume}{18}, \bibinfo{number}{2} (\bibinfo{year}{2022}), \bibinfo{pages}{1--20}.
\newblock


\bibitem[Sun et~al\mbox{.}(2023)]%
        {ref:llm_rec_2}
\bibfield{author}{\bibinfo{person}{Weiwei Sun}, \bibinfo{person}{Lingyong Yan}, \bibinfo{person}{Xinyu Ma}, \bibinfo{person}{Shuaiqiang Wang}, \bibinfo{person}{Pengjie Ren}, \bibinfo{person}{Zhumin Chen}, \bibinfo{person}{Dawei Yin}, {and} \bibinfo{person}{Zhaochun Ren}.} \bibinfo{year}{2023}\natexlab{}.
\newblock \showarticletitle{Is ChatGPT Good at Search? Investigating Large Language Models as Re-Ranking Agents}.
\newblock  (\bibinfo{year}{2023}), \bibinfo{pages}{14918--14937}.
\newblock


\bibitem[Sun et~al\mbox{.}(2024)]%
        {sun2024parrot}
\bibfield{author}{\bibinfo{person}{Yuchong Sun}, \bibinfo{person}{Che Liu}, \bibinfo{person}{Kun Zhou}, \bibinfo{person}{Jinwen Huang}, \bibinfo{person}{Ruihua Song}, \bibinfo{person}{Wayne~Xin Zhao}, \bibinfo{person}{Fuzheng Zhang}, \bibinfo{person}{Di Zhang}, {and} \bibinfo{person}{Kun Gai}.} \bibinfo{year}{2024}\natexlab{}.
\newblock \showarticletitle{Parrot: Enhancing multi-turn instruction following for large language models}. In \bibinfo{booktitle}{\emph{Proceedings of the 62nd Annual Meeting of the Association for Computational Linguistics (Volume 1: Long Papers)}}. \bibinfo{pages}{9729--9750}.
\newblock


\bibitem[Tang and Wang(2018)]%
        {ref:caser}
\bibfield{author}{\bibinfo{person}{Jiaxi Tang} {and} \bibinfo{person}{Ke Wang}.} \bibinfo{year}{2018}\natexlab{}.
\newblock \showarticletitle{Personalized top-n sequential recommendation via convolutional sequence embedding}. In \bibinfo{booktitle}{\emph{Proceedings of the eleventh ACM international conference on web search and data mining}}. \bibinfo{pages}{565--573}.
\newblock


\bibitem[Touvron et~al\mbox{.}(2023)]%
        {ref:llama}
\bibfield{author}{\bibinfo{person}{Hugo Touvron}, \bibinfo{person}{Thibaut Lavril}, \bibinfo{person}{Gautier Izacard}, \bibinfo{person}{Xavier Martinet}, \bibinfo{person}{Marie{-}Anne Lachaux}, \bibinfo{person}{Timoth{\'{e}}e Lacroix}, \bibinfo{person}{Baptiste Rozi{\`{e}}re}, \bibinfo{person}{Naman Goyal}, \bibinfo{person}{Eric Hambro}, \bibinfo{person}{Faisal Azhar}, \bibinfo{person}{Aur{\'{e}}lien Rodriguez}, \bibinfo{person}{Armand Joulin}, \bibinfo{person}{Edouard Grave}, {and} \bibinfo{person}{Guillaume Lample}.} \bibinfo{year}{2023}\natexlab{}.
\newblock \showarticletitle{LLaMA: Open and Efficient Foundation Language Models}.
\newblock \bibinfo{journal}{\emph{CoRR}}  \bibinfo{volume}{abs/2302.13971} (\bibinfo{year}{2023}).
\newblock
\href{https://doi.org/10.48550/ARXIV.2302.13971}{doi:\nolinkurl{10.48550/ARXIV.2302.13971}}
\showeprint[arXiv]{2302.13971}


\bibitem[Wang et~al\mbox{.}(2024)]%
        {wang2024causal}
\bibfield{author}{\bibinfo{person}{Jiawei Wang}, \bibinfo{person}{Da Cao}, \bibinfo{person}{Shaofei Lu}, \bibinfo{person}{Zhanchang Ma}, \bibinfo{person}{Junbin Xiao}, {and} \bibinfo{person}{Tat-Seng Chua}.} \bibinfo{year}{2024}\natexlab{}.
\newblock \showarticletitle{Causal-driven Large Language Models with Faithful Reasoning for Knowledge Question Answering}. In \bibinfo{booktitle}{\emph{Proceedings of the 32nd ACM International Conference on Multimedia}}. \bibinfo{pages}{4331--4340}.
\newblock


\bibitem[Wu et~al\mbox{.}(2023)]%
        {ref:survey_llm4rec}
\bibfield{author}{\bibinfo{person}{Likang Wu}, \bibinfo{person}{Zhi Zheng}, \bibinfo{person}{Zhaopeng Qiu}, \bibinfo{person}{Hao Wang}, \bibinfo{person}{Hongchao Gu}, \bibinfo{person}{Tingjia Shen}, \bibinfo{person}{Chuan Qin}, \bibinfo{person}{Chen Zhu}, \bibinfo{person}{Hengshu Zhu}, \bibinfo{person}{Qi Liu}, {et~al\mbox{.}}} \bibinfo{year}{2023}\natexlab{}.
\newblock \showarticletitle{A Survey on Large Language Models for Recommendation}.
\newblock \bibinfo{journal}{\emph{arXiv preprint arXiv:2305.19860}} (\bibinfo{year}{2023}).
\newblock


\bibitem[Wu et~al\mbox{.}(2024b)]%
        {ref:llm_survey}
\bibfield{author}{\bibinfo{person}{Likang Wu}, \bibinfo{person}{Zhi Zheng}, \bibinfo{person}{Zhaopeng Qiu}, \bibinfo{person}{Hao Wang}, \bibinfo{person}{Hongchao Gu}, \bibinfo{person}{Tingjia Shen}, \bibinfo{person}{Chuan Qin}, \bibinfo{person}{Chen Zhu}, \bibinfo{person}{Hengshu Zhu}, \bibinfo{person}{Qi Liu}, \bibinfo{person}{Hui Xiong}, {and} \bibinfo{person}{Enhong Chen}.} \bibinfo{year}{2024}\natexlab{b}.
\newblock \showarticletitle{A survey on large language models for recommendation}.
\newblock \bibinfo{journal}{\emph{World Wide Web {(WWW)}}} \bibinfo{volume}{27}, \bibinfo{number}{5} (\bibinfo{year}{2024}), \bibinfo{pages}{60}.
\newblock


\bibitem[Wu et~al\mbox{.}(2024a)]%
        {wu2024nextgpt}
\bibfield{author}{\bibinfo{person}{Shengqiong Wu}, \bibinfo{person}{Hao Fei}, \bibinfo{person}{Leigang Qu}, \bibinfo{person}{Wei Ji}, {and} \bibinfo{person}{Tat{-}Seng Chua}.} \bibinfo{year}{2024}\natexlab{a}.
\newblock \showarticletitle{NExT-GPT: Any-to-Any Multimodal {LLM}}. In \bibinfo{booktitle}{\emph{{ICML}}}. \bibinfo{publisher}{OpenReview.net}.
\newblock


\bibitem[Xu et~al\mbox{.}(2024)]%
        {xu2024fairrec}
\bibfield{author}{\bibinfo{person}{Chen Xu}, \bibinfo{person}{Wenjie Wang}, \bibinfo{person}{Yuxin Li}, \bibinfo{person}{Liang Pang}, \bibinfo{person}{Jun Xu}, {and} \bibinfo{person}{Tat{-}Seng Chua}.} \bibinfo{year}{2024}\natexlab{}.
\newblock \showarticletitle{A Study of Implicit Ranking Unfairness in Large Language Models}. In \bibinfo{booktitle}{\emph{{EMNLP} (Findings)}}. \bibinfo{publisher}{Association for Computational Linguistics}, \bibinfo{pages}{7957--7970}.
\newblock


\bibitem[Yao et~al\mbox{.}(2022)]%
        {ref:device_cloud}
\bibfield{author}{\bibinfo{person}{Jiangchao Yao}, \bibinfo{person}{Feng Wang}, \bibinfo{person}{Xichen Ding}, \bibinfo{person}{Shaohu Chen}, \bibinfo{person}{Bo Han}, \bibinfo{person}{Jingren Zhou}, {and} \bibinfo{person}{Hongxia Yang}.} \bibinfo{year}{2022}\natexlab{}.
\newblock \showarticletitle{Device-cloud Collaborative Recommendation via Meta Controller}. In \bibinfo{booktitle}{\emph{{KDD}}}. \bibinfo{publisher}{{ACM}}, \bibinfo{pages}{4353--4362}.
\newblock


\bibitem[Yao et~al\mbox{.}(2021)]%
        {ref:device_cloud_dccl}
\bibfield{author}{\bibinfo{person}{Jiangchao Yao}, \bibinfo{person}{Feng Wang}, \bibinfo{person}{Kunyang Jia}, \bibinfo{person}{Bo Han}, \bibinfo{person}{Jingren Zhou}, {and} \bibinfo{person}{Hongxia Yang}.} \bibinfo{year}{2021}\natexlab{}.
\newblock \showarticletitle{Device-Cloud Collaborative Learning for Recommendation}. In \bibinfo{booktitle}{\emph{{KDD}}}. \bibinfo{publisher}{{ACM}}, \bibinfo{pages}{3865--3874}.
\newblock


\bibitem[Zhang et~al\mbox{.}(2024a)]%
        {zhang2024transfr}
\bibfield{author}{\bibinfo{person}{Honglei Zhang}, \bibinfo{person}{He Liu}, \bibinfo{person}{Haoxuan Li}, {and} \bibinfo{person}{Yidong Li}.} \bibinfo{year}{2024}\natexlab{a}.
\newblock \showarticletitle{Transfr: Transferable federated recommendation with pre-trained language models}.
\newblock \bibinfo{journal}{\emph{arXiv:2402.01124}} (\bibinfo{year}{2024}).
\newblock


\bibitem[Zhang et~al\mbox{.}(2023b)]%
        {zhang2023mining}
\bibfield{author}{\bibinfo{person}{Jinghao Zhang}, \bibinfo{person}{Qiang Liu}, \bibinfo{person}{Shu Wu}, {and} \bibinfo{person}{Liang Wang}.} \bibinfo{year}{2023}\natexlab{b}.
\newblock \showarticletitle{Mining Stable Preferences: Adaptive Modality Decorrelation for Multimedia Recommendation}. In \bibinfo{booktitle}{\emph{Proceedings of the 46th International ACM SIGIR Conference on Research and Development in Information Retrieval}}. \bibinfo{pages}{443--452}.
\newblock


\bibitem[Zhang et~al\mbox{.}(2023c)]%
        {ref:instructrec}
\bibfield{author}{\bibinfo{person}{Junjie Zhang}, \bibinfo{person}{Ruobing Xie}, \bibinfo{person}{Yupeng Hou}, \bibinfo{person}{Wayne~Xin Zhao}, \bibinfo{person}{Leyu Lin}, {and} \bibinfo{person}{Ji{-}Rong Wen}.} \bibinfo{year}{2023}\natexlab{c}.
\newblock \showarticletitle{Recommendation as Instruction Following: {A} Large Language Model Empowered Recommendation Approach}.
\newblock \bibinfo{journal}{\emph{CoRR}}  \bibinfo{volume}{abs/2305.07001} (\bibinfo{year}{2023}).
\newblock
\href{https://doi.org/10.48550/ARXIV.2305.07001}{doi:\nolinkurl{10.48550/ARXIV.2305.07001}}
\showeprint[arXiv]{2305.07001}


\bibitem[Zhang et~al\mbox{.}(2023d)]%
        {ref:llm4rec_understand_user_query}
\bibfield{author}{\bibinfo{person}{Junjie Zhang}, \bibinfo{person}{Ruobing Xie}, \bibinfo{person}{Yupeng Hou}, \bibinfo{person}{Wayne~Xin Zhao}, \bibinfo{person}{Leyu Lin}, {and} \bibinfo{person}{Ji{-}Rong Wen}.} \bibinfo{year}{2023}\natexlab{d}.
\newblock \showarticletitle{Recommendation as Instruction Following: {A} Large Language Model Empowered Recommendation Approach}.
\newblock \bibinfo{journal}{\emph{CoRR}}  \bibinfo{volume}{abs/2305.07001} (\bibinfo{year}{2023}).
\newblock


\bibitem[Zhang et~al\mbox{.}(2021)]%
        {zhang2021mining}
\bibfield{author}{\bibinfo{person}{Jinghao Zhang}, \bibinfo{person}{Yanqiao Zhu}, \bibinfo{person}{Qiang Liu}, \bibinfo{person}{Shu Wu}, \bibinfo{person}{Shuhui Wang}, {and} \bibinfo{person}{Liang Wang}.} \bibinfo{year}{2021}\natexlab{}.
\newblock \showarticletitle{Mining latent structures for multimedia recommendation}. In \bibinfo{booktitle}{\emph{Proceedings of the 29th ACM international conference on multimedia}}. \bibinfo{pages}{3872--3880}.
\newblock


\bibitem[Zhang et~al\mbox{.}(2022)]%
        {zhang2022latent}
\bibfield{author}{\bibinfo{person}{Jinghao Zhang}, \bibinfo{person}{Yanqiao Zhu}, \bibinfo{person}{Qiang Liu}, \bibinfo{person}{Mengqi Zhang}, \bibinfo{person}{Shu Wu}, {and} \bibinfo{person}{Liang Wang}.} \bibinfo{year}{2022}\natexlab{}.
\newblock \showarticletitle{Latent structure mining with contrastive modality fusion for multimedia recommendation}.
\newblock \bibinfo{journal}{\emph{IEEE Transactions on Knowledge and Data Engineering}} \bibinfo{volume}{35}, \bibinfo{number}{9} (\bibinfo{year}{2022}), \bibinfo{pages}{9154--9167}.
\newblock


\bibitem[Zhang et~al\mbox{.}(2024c)]%
        {zhang2024saqrec}
\bibfield{author}{\bibinfo{person}{Kepu Zhang}, \bibinfo{person}{Teng Shi}, \bibinfo{person}{Sunhao Dai}, \bibinfo{person}{Xiao Zhang}, \bibinfo{person}{Yinfeng Li}, \bibinfo{person}{Jing Lu}, \bibinfo{person}{Xiaoxue Zang}, \bibinfo{person}{Yang Song}, {and} \bibinfo{person}{Jun Xu}.} \bibinfo{year}{2024}\natexlab{c}.
\newblock \showarticletitle{SAQRec: Aligning Recommender Systems to User Satisfaction via Questionnaire Feedback}. In \bibinfo{booktitle}{\emph{{CIKM}}}. \bibinfo{publisher}{{ACM}}, \bibinfo{pages}{3165--3175}.
\newblock


\bibitem[Zhang et~al\mbox{.}(2024b)]%
        {zhang2024llasa}
\bibfield{author}{\bibinfo{person}{Shuo Zhang}, \bibinfo{person}{Boci Peng}, \bibinfo{person}{Xinping Zhao}, \bibinfo{person}{Boren Hu}, \bibinfo{person}{Yun Zhu}, \bibinfo{person}{Yanjia Zeng}, {and} \bibinfo{person}{Xuming Hu}.} \bibinfo{year}{2024}\natexlab{b}.
\newblock \showarticletitle{LLaSA: Large Language and E-Commerce Shopping Assistant}. In \bibinfo{booktitle}{\emph{Amazon KDD Cup 2024 Workshop}}.
\newblock


\bibitem[Zhang et~al\mbox{.}(2023a)]%
        {zhang2023collm}
\bibfield{author}{\bibinfo{person}{Yang Zhang}, \bibinfo{person}{Fuli Feng}, \bibinfo{person}{Jizhi Zhang}, \bibinfo{person}{Keqin Bao}, \bibinfo{person}{Qifan Wang}, {and} \bibinfo{person}{Xiangnan He}.} \bibinfo{year}{2023}\natexlab{a}.
\newblock \showarticletitle{Collm: Integrating collaborative embeddings into large language models for recommendation}.
\newblock \bibinfo{journal}{\emph{arXiv preprint arXiv:2310.19488}} (\bibinfo{year}{2023}).
\newblock


\bibitem[Zhao et~al\mbox{.}(2023)]%
        {zhao2023popularity}
\bibfield{author}{\bibinfo{person}{Jujia Zhao}, \bibinfo{person}{Wenjie Wang}, \bibinfo{person}{Xinyu Lin}, \bibinfo{person}{Leigang Qu}, \bibinfo{person}{Jizhi Zhang}, {and} \bibinfo{person}{Tat{-}Seng Chua}.} \bibinfo{year}{2023}\natexlab{}.
\newblock \showarticletitle{Popularity-aware Distributionally Robust Optimization for Recommendation System}. In \bibinfo{booktitle}{\emph{{CIKM}}}. \bibinfo{publisher}{{ACM}}, \bibinfo{pages}{4967--4973}.
\newblock


\bibitem[Zhao et~al\mbox{.}(2024c)]%
        {zhao2024denoising}
\bibfield{author}{\bibinfo{person}{Jujia Zhao}, \bibinfo{person}{Wang Wenjie}, \bibinfo{person}{Yiyan Xu}, \bibinfo{person}{Teng Sun}, \bibinfo{person}{Fuli Feng}, {and} \bibinfo{person}{Tat-Seng Chua}.} \bibinfo{year}{2024}\natexlab{c}.
\newblock \showarticletitle{Denoising diffusion recommender model}. In \bibinfo{booktitle}{\emph{Proceedings of the 47th International ACM SIGIR Conference on Research and Development in Information Retrieval}}. \bibinfo{pages}{1370--1379}.
\newblock


\bibitem[Zhao et~al\mbox{.}(2024a)]%
        {ref:llm_rec1}
\bibfield{author}{\bibinfo{person}{Zihuai Zhao}, \bibinfo{person}{Wenqi Fan}, \bibinfo{person}{Jiatong Li}, \bibinfo{person}{Yunqing Liu}, \bibinfo{person}{Xiaowei Mei}, \bibinfo{person}{Yiqi Wang}, \bibinfo{person}{Zhen Wen}, \bibinfo{person}{Fei Wang}, \bibinfo{person}{Xiangyu Zhao}, \bibinfo{person}{Jiliang Tang}, {and} \bibinfo{person}{Qing Li}.} \bibinfo{year}{2024}\natexlab{a}.
\newblock \showarticletitle{Recommender Systems in the Era of Large Language Models (LLMs)}.
\newblock \bibinfo{journal}{\emph{{IEEE} Trans. Knowl. Data Eng.}} \bibinfo{volume}{36}, \bibinfo{number}{11} (\bibinfo{year}{2024}), \bibinfo{pages}{6889--6907}.
\newblock


\bibitem[Zhao et~al\mbox{.}(2024b)]%
        {ref:llm_rec_4}
\bibfield{author}{\bibinfo{person}{Zihuai Zhao}, \bibinfo{person}{Wenqi Fan}, \bibinfo{person}{Jiatong Li}, \bibinfo{person}{Yunqing Liu}, \bibinfo{person}{Xiaowei Mei}, \bibinfo{person}{Yiqi Wang}, \bibinfo{person}{Zhen Wen}, \bibinfo{person}{Fei Wang}, \bibinfo{person}{Xiangyu Zhao}, \bibinfo{person}{Jiliang Tang}, {and} \bibinfo{person}{Qing Li}.} \bibinfo{year}{2024}\natexlab{b}.
\newblock \showarticletitle{Recommender Systems in the Era of Large Language Models (LLMs)}.
\newblock \bibinfo{journal}{\emph{{IEEE} Trans. Knowl. Data Eng.}} \bibinfo{volume}{36}, \bibinfo{number}{11} (\bibinfo{year}{2024}), \bibinfo{pages}{6889--6907}.
\newblock


\bibitem[Zheng et~al\mbox{.}(2022)]%
        {zheng2022ddghm}
\bibfield{author}{\bibinfo{person}{Xiaolin Zheng}, \bibinfo{person}{Jiajie Su}, \bibinfo{person}{Weiming Liu}, {and} \bibinfo{person}{Chaochao Chen}.} \bibinfo{year}{2022}\natexlab{}.
\newblock \showarticletitle{DDGHM: Dual dynamic graph with hybrid metric training for cross-domain sequential recommendation}. In \bibinfo{booktitle}{\emph{Proceedings of the 30th ACM International Conference on Multimedia}}. \bibinfo{pages}{471--481}.
\newblock


\bibitem[Zhou et~al\mbox{.}(2018)]%
        {ref:din}
\bibfield{author}{\bibinfo{person}{Guorui Zhou}, \bibinfo{person}{Xiaoqiang Zhu}, \bibinfo{person}{Chengru Song}, \bibinfo{person}{Ying Fan}, \bibinfo{person}{Han Zhu}, \bibinfo{person}{Xiao Ma}, \bibinfo{person}{Yanghui Yan}, \bibinfo{person}{Junqi Jin}, \bibinfo{person}{Han Li}, {and} \bibinfo{person}{Kun Gai}.} \bibinfo{year}{2018}\natexlab{}.
\newblock \showarticletitle{Deep Interest Network for Click-Through Rate Prediction}. In \bibinfo{booktitle}{\emph{{KDD}}}. \bibinfo{publisher}{{ACM}}, \bibinfo{pages}{1059--1068}.
\newblock


\bibitem[Zhu et~al\mbox{.}(2024a)]%
        {zhu2024model}
\bibfield{author}{\bibinfo{person}{Didi Zhu}, \bibinfo{person}{Zhongyisun Sun}, \bibinfo{person}{Zexi Li}, \bibinfo{person}{Tao Shen}, \bibinfo{person}{Ke Yan}, \bibinfo{person}{Shouhong Ding}, \bibinfo{person}{Chao Wu}, {and} \bibinfo{person}{Kun Kuang}.} \bibinfo{year}{2024}\natexlab{a}.
\newblock \showarticletitle{Model Tailor: Mitigating Catastrophic Forgetting in Multi-modal Large Language Models}. In \bibinfo{booktitle}{\emph{Forty-first International Conference on Machine Learning}}.
\newblock


\bibitem[Zhu et~al\mbox{.}(2024b)]%
        {zhu2024efficient}
\bibfield{author}{\bibinfo{person}{Yun Zhu}, \bibinfo{person}{Yaoke Wang}, \bibinfo{person}{Haizhou Shi}, {and} \bibinfo{person}{Siliang Tang}.} \bibinfo{year}{2024}\natexlab{b}.
\newblock \showarticletitle{Efficient Tuning and Inference for Large Language Models on Textual Graphs}. In \bibinfo{booktitle}{\emph{Proceedings of the Thirty-Third International Joint Conference on Artificial Intelligence, {IJCAI-24}}}, \bibfield{editor}{\bibinfo{person}{Kate Larson}} (Ed.). \bibinfo{publisher}{International Joint Conferences on Artificial Intelligence Organization}, \bibinfo{pages}{5734--5742}.
\newblock
\href{https://doi.org/10.24963/ijcai.2024/634}{doi:\nolinkurl{10.24963/ijcai.2024/634}}
\newblock
\shownote{Main Track}.


\end{thebibliography}
\clearpage
\appendix

\section{Appendix}
\label{sec:appendix}
This is the Appendix for ``Collaboration of Large Language Models and Small Recommendation Models for Device-Cloud Recommendation''.

\subsection{Supplementary Methodology}

\subsubsection{Pseudo code of LSCRec}
\label{sec:pseudo_code}

Algorithm~\ref{alg:pseudo_code} shows the pseudo code of LSCRec. $(x)$ represents that $x$ is a intermediate variable.
\begin{algorithm}[!ht]
\begin{flushleft}
  \caption{Pseudo Code of LSC4Rec}
    
\textbf{Strategy 1:}~\colorbox{gray!30}{$\rhd$~\emph{Collaborative Training}}
    \resizebox{0.46\textwidth}{!}{
    \begin{tcolorbox}[sharp corners, colframe=gray!80!white, colback=white, boxrule=0.5mm, left=0pt, right=0pt, top=0pt, bottom=0pt, boxsep=5pt]
    \begin{algorithmic}
    \State \textbf{Target}: Pretrained LLM $\mathcal{M}_L$, Pretrained SRM $\mathcal{M}_S$ $\mapsto$ Updated SRM $\mathcal{M}_S$
    \State \textbf{Input}: Historical data $\mathcal{D}_H$
    \State \textbf{Output}: Prediction $\hat{Y}$
    \end{algorithmic}
    \end{tcolorbox}
    }

\textbf{Strategy 2:}~\colorbox{gray!30}{$\rhd$~\emph{Collaborative Inference}}
\resizebox{0.46\textwidth}{!}{
    \begin{tcolorbox}[sharp corners, colframe=gray!80!white, colback=white, boxrule=0.5mm, left=0pt, right=0pt, top=0pt, bottom=0pt, boxsep=5pt]
    \begin{algorithmic}
    \State \textbf{Target}: Real-time data $\mathcal{D}_R$ $\mapsto$ Prediction $\hat{Y}$
    \State \textbf{Input}: Real-time data $\mathcal{D}_R$
    \State \textbf{Output}: (Candidate Item Set $\mathcal{S}$), (Initial Ranking $\hat{Y}_{\rm{init}}$ generated by $\mathcal{M}_L$), (Re-Ranking $\hat{Y}_{\rm{rerank}}$ generated by $\mathcal{M}_S$), Prediction $\hat{Y}$
    \end{algorithmic}
    \end{tcolorbox}
    }

\textbf{Strategy 3:}~\colorbox{gray!30}{$\rhd$~\emph{Collaborative-decision Request}}
\resizebox{0.46\textwidth}{!}{
    \begin{tcolorbox}[sharp corners, colframe=gray!80!white, colback=white, boxrule=0.5mm, left=0pt, right=0pt, top=0pt, bottom=0pt, boxsep=5pt]
    \begin{algorithmic}
    \State \textbf{Target}: Initial Ranking $\hat{Y}_{\rm{init}}$, Re-Ranking $\hat{Y}_{\rm{rerank}}$ $\mapsto$ Inconsistency $c$
    \State \textbf{Input}: Initial Ranking $\hat{Y}_{\rm{init}}$, Re-Ranking $\hat{Y}_{\rm{rerank}}$
    \State \textbf{Output}: Inconsistency $c$
    \end{algorithmic}
    \end{tcolorbox}
    }

\textbf{Overview:}~\colorbox{gray!30}{$\rhd$~\emph{Training Procedure}}\\
    \textbf{Input}: Historical data $\mathcal{D}_H$, Real-time data $\mathcal{D}_R$ \\
    \textbf{Output}: Prediction $\hat{y}$.\\
    \textbf{Initialization}: Randomly initialize the $\mathcal{M}_L$ and $\mathcal{M}_S$ \\
    \Repeat {Convergence}{
        \If{$\mathcal{M}_L$ and $\mathcal{M}_S$ have not yet been well-trained} 
        {
           Calculate loss as follows (see Eq.\ref{eq:loss_general} for the details),\\
           $\mathcal{L}=\sum_{u, v, s, y\in\mathcal{D}_H}l(y, \hat{y}=\mathcal{M}(u, v, s))$. \\
        }
    }
    \Repeat {Convergence}{
        \If{$\mathcal{M}_S$ has not yet been well-trained} 
        {
            Calculate loss as follows (see Eq.\ref{eq:candidate_cotraining}$\sim$\ref{eq:loss_srm_realtime_train} for the details),\\
            $\mathcal{L}=\sum\nolimits_{u, v, s, y\in\mathcal{D}_H}l(y, \hat{y}=\mathcal{M}_S(u, v, s)|S_{Aug}).$\\
        }
    }
    \Return{$\mathcal{M}_L$, $\mathcal{M}_S$}.\\
\textbf{Overview:}~\colorbox{gray!30}{$\rhd$~\emph{Inference Procedure}}\\
\textbf{Input}: Real-time data $\mathcal{D}_R$\\
\textbf{Output}: Prediction $\hat{Y}$, Inconsistency $c$\\
$S, \hat{Y}_{\rm{init}}$ predicted by LLM, $\hat{Y}_{\rm{rerank}}$ predicted by SRM based on $S$.\\
$c$ calculated based on $\hat{Y}_{\rm{init}}$ and $\hat{Y}_{\rm{rerank}}$ to help decide whether to request new $S, \hat{Y}_{\rm{init}}$ from LLM.
\label{alg:pseudo_code}
\end{flushleft}
\end{algorithm}

\subsection{Supplementary Experiments}
\subsubsection{Datasets.}
The statistics of the datasets used in the experiments is shown in Table~\ref{tab:statistics_of_datasets}.

\begin{table}[ht]
\caption{Statistics of Datasets.} 
\label{tab:statistics_of_datasets}
\vspace{-0.3cm}
\resizebox{0.3\textwidth}{!}{
\begin{tabular}{c|c|c|c}
\toprule[2pt]
\textbf{Dataset} & \textbf{Beauty} & \textbf{Toys} & \textbf{Yelp} \\
\midrule \midrule
\#User & 22,363 & 19,412 & 30,431 \\
\rowcolor[HTML]{F2F2F2} 
\#Item & 12,101 & 11,924 & 20,033 \\
\#interaction & 198,502 & 167,597 & 316,354 \\
\rowcolor[HTML]{F2F2F2} 
Density & 0.00073352 & 0.00072406 & 0.00051893 \\
\bottomrule[2pt]
\end{tabular}
}
\vspace{-0.2cm}
\end{table}

\subsubsection{Hyperparameters and Training Schedules}
\label{sec:appendix_implementation_detail}
We summarize the hyperparameters and training schedules of the LLM and SRM used in the experiments. 
Table~\ref{tab:hyperparameters_and_training_schedule} shows the settings of the SRM training. Table~\ref{tab:hyperparameters_and_training_schedule_small} shows the setting of the collaborative training, collaborative inference, and collaborative-decision request.


\begin{table}[!ht]
    \caption{Hyperparameters of LLM training.}
    \label{tab:hyperparameters_and_training_schedule_large}
    \centering
    \vspace{-0.3cm}
    \setlength{\arrayrulewidth}{0.5pt}
 \resizebox{0.44\textwidth}{!}{
    \begin{tabular}{c|c|c|c}
    \toprule[2pt]
    \textbf{Dataset} & \textbf{LLM} & \textbf{Hyperparameter} & \textbf{Setting} \\ 
    \midrule
    \midrule
    \multirow{10}{*}{\makecell[c]{Beauty\\Toys\\Yelp}} & \multirow{10}{*}{\makecell[c]{P5}} & GPU & Tesla A100 (40GB) \\ 
    & & \cellcolor{gray!15}Batch Size & \cellcolor{gray!15}32 \\ 
    & & Max Text Length & 512 \\ 
    & & \cellcolor{gray!15}Gen Max Length & \cellcolor{gray!15}64 \\
    & & Learning Rate & 1e-3 \\ 
    & & \cellcolor{gray!15}Weight Decay & \cellcolor{gray!15}0.05(Beauty, Toys), 0.02(Yelp) \\ 
    & & Optimizer & Adamw \\
    & & \cellcolor{gray!15}Task & \cellcolor{gray!15}Sequential \\ 
    & & Negative Sampling Rate & 1:99 \\
    & & \cellcolor{gray!15}Beam Search & \cellcolor{gray!15}20/50 \\ 
     \toprule
     \toprule
    \multirow{8}{*}{\makecell[c]{Beauty\\Toys\\Yelp}} & \multirow{8}{*}{\makecell[c]{POD}} & GPU & Tesla A100 (40GB) \\ 
    & & \cellcolor{gray!15}Batch Size & \cellcolor{gray!15}64 \\ 
    & & Learning Rate & 5e-4 \\ 
    & & \cellcolor{gray!15}Weight Decay & \cellcolor{gray!15}1e-2 \\ 
    & & Optimizer & Adamw \\
    & & \cellcolor{gray!15}Task & \cellcolor{gray!15}Sequential \\ 
    & & Negative Sampling Rate & 1:99 \\
    & & \cellcolor{gray!15}Beam Search & \cellcolor{gray!15}20/50 \\ 
     \bottomrule[2pt]
     
    \end{tabular}
   }
    \vspace{-0.3cm}
\end{table}

\begin{table}[!h]
    \caption{Hyperparameters of SRM training.}
    \label{tab:hyperparameters_and_training_schedule}
    \centering
    \vspace{-0.3cm}
    \setlength{\arrayrulewidth}{0.5pt}
 \resizebox{0.42\textwidth}{!}{
    \begin{tabular}{c|c|c|c}
    \toprule[2pt]
    \textbf{Dataset} & \textbf{SRM} & \textbf{Hyperparameter} & \textbf{Setting} \\ 
    \midrule
    \midrule
    \multirow{7}{*}{\makecell[c]{Beauty\\Toys\\Yelp}} & \multirow{7}{*}{\makecell[c]{DIN\\GRU4Rec\\SASRec}} & GPU & Tesla A100 (40GB) \\ 
    \multirow{7}{*}{} & & \cellcolor{gray!15}Optimizer & \cellcolor{gray!15}Adam\\ 
    \multirow{7}{*}{} & & \makecell[c]{Learning rate} & 1e-3\\ 
    \multirow{7}{*}{} & & \cellcolor{gray!15}{Batch size} & \cellcolor{gray!15}1024 \\ 
    \multirow{7}{*}{} & & {Sequence length} & 10 \\ 
    \multirow{7}{*}{} & & \cellcolor{gray!15}\makecell[c]{the Dimension of Embedding} & \cellcolor{gray!15}1×32 \\  
    \multirow{7}{*}{} & & \makecell[c]{the Amount of MLP} & 2 \\  
    \bottomrule[2pt]
    \end{tabular}
   }
    \vspace{-0.3cm}
\end{table}

\begin{table}[!h]
    \caption{Hyperparameters of collaborative training, collaborative inference, and collaborative-decision request.}
    \label{tab:hyperparameters_and_training_schedule_small}
    \centering
    \vspace{-0.3cm}
    \setlength{\arrayrulewidth}{0.5pt}
 \resizebox{0.43\textwidth}{!}{
    \begin{tabular}{c|c|c|c|c}
    \toprule[2pt]
    \textbf{Dataset} & \textbf{LLM} & \textbf{SRM} & \textbf{Hyperparameter} & \textbf{Setting} \\
    \midrule
    \midrule
    \multirow{7}{*}{\makecell[c]{Beauty\\Toys\\Yelp}}
    & \multirow{7}{*}{\makecell[c]{P5\\POD}}
    & \multirow{7}{*}{\makecell[c]{DIN\\GRU4Rec\\SASRec}}
    & Length of Candidate list  & 20/50 \\ 
    & & & \cellcolor{gray!15}Valid user behavior length & \cellcolor{gray!15}Dynamic \\ 
    & & &  User behavior length & 10 \\ 
    & & &  \cellcolor{gray!15}Sequence length & \cellcolor{gray!15}50 \\ 
    & & &  Learning Rate & 1e-3 \\
    & & &  \cellcolor{gray!15}Dimension of embedding & \cellcolor{gray!15}1×32 \\ 
    & & &  Optimizer & Adam \\ 
    \bottomrule[2pt]
    \end{tabular}
   }
    \vspace{-0.3cm}
\end{table}

\end{document}